\documentclass[a4paper,12pt]{article}
\usepackage{amsfonts,amsthm,amsmath,amssymb,graphicx}
\numberwithin{equation}{section}
\begin{document}

\newtheorem{definition}{Definition}[section]
\newcommand{\be}{\begin{equation}}
\newcommand{\ee}{\end{equation}}
\newcommand{\bea}{\begin{eqnarray}}
\newcommand{\eea}{\end{eqnarray}}
\newcommand{\LE}{\left[}
\newcommand{\R}{\right]}
\newcommand{\nn}{\nonumber}
\newcommand{\Tr}{\text{Tr}}
\newcommand{\N}{\mathcal{N}}
\newcommand{\G}{\Gamma}
\newcommand{\vf}{\varphi}
\newcommand{\LL}{\mathcal{L}}
\newcommand{\Op}{\mathcal{O}}
\newcommand{\HH}{\mathcal{H}}
\newcommand{\arctanh}{\text{arctanh}}
\newcommand{\up}{\uparrow}
\newcommand{\down}{\downarrow}
\newcommand{\ket}[1]{\left| #1 \right>}
\newcommand{\bra}[1]{\left< #1 \right|}
\newcommand{\ketbra}[1]{\left|#1\right>\left<#1\right|}
\newcommand{\rd}{\partial}
\newcommand{\de}{\partial}
\newcommand{\ba}{\begin{eqnarray}}
\newcommand{\ea}{\end{eqnarray}}
\newcommand{\db}{\bar{\partial}}
\newcommand{\we}{\wedge}
\newcommand{\ca}{\mathcal}
\newcommand{\lr}{\leftrightarrow}
\newcommand{\f}{\frac}
\newcommand{\s}{\sqrt}
\newcommand{\vp}{\varphi}
\newcommand{\hvp}{\hat{\varphi}}
\newcommand{\tvp}{\tilde{\varphi}}
\newcommand{\tp}{\tilde{\phi}}
\newcommand{\ti}{\tilde}
\newcommand{\ap}{\alpha}
\newcommand{\pr}{\propto}
\newcommand{\mb}{\mathbf}
\newcommand{\ddd}{\cdot\cdot\cdot}
\newcommand{\no}{\nonumber \\}
\newcommand{\la}{\langle}
\newcommand{\lb}{\rangle}
\newcommand{\ep}{\epsilon}
 \def\we{\wedge}
 \def\lr{\leftrightarrow}
 \def\f {\frac}
 \def\ti{\tilde}
 \def\ap{\alpha}
 \def\pr{\propto}
 \def\mb{\mathbf}
 \def\ddd{\cdot\cdot\cdot}
 \def\no{\nonumber \\}
 \def\la{\langle}
 \def\lb{\rangle}
 \def\ep{\epsilon}
\newcommand{\mcl}{\mathcal}
 \def\g{\gamma}
\def\tr{\text{tr}}

\begin{titlepage}
\thispagestyle{empty}

\begin{flushright}
NORDITA-2015-137\\
EFI-15-38\\
YITP-15-118\\
\end{flushright}
\bigskip

\begin{center}
\noindent{\large \textbf{Charged Entanglement Entropy of Local Operators }}\\
\vspace{2cm}

Pawe{\l} Caputa $^{a}$, Masahiro Nozaki $^{b}$ and Tokiro Numasawa $^{c}$ \\

\vspace{1cm}
{\it $^{a}$Nordita, Stockholm University,\\
Roslagstullsbacken 23, SE-106 91 Stockholm, Sweden\\}

{\it $^{b}$Kadanoff Center for Theoretical Physics, University of Chicago,\\
Chicago, Illinois 60637, USA \\}

{\it
 $^{c}$Yukawa Institute for Theoretical Physics,
Kyoto University,\\ Kyoto 606-8502, Japan\\}

\vskip 4em
\end{center}
\begin{abstract}
\vskip 1em
In this work we consider the time evolution of charged Renyi entanglement entropies after exciting the vacuum with local fermionic operators. In order to explore the information contained in charged Renyi entropies, we perform computations of their excess due to the operator excitation in 2d CFT, free fermionic field theories in various dimensions as well as holographically. In the analysis we focus on the dependence on the entanglement charge, the chemical potential and the spacetime dimension. We find that excesses of charged (Renyi) entanglement entropy can be interpreted in terms of charged quasiparticles. Moreover, we show that by appropriately tuning the chemical potential, charged Renyi entropies can be used to extract entanglement in a certain charge sector of the excited state.   
\end{abstract}
\end{titlepage}

\tableofcontents

\section{Introduction and Summary}
Measures of entanglement in field theories provide a zooming glass that can detect new phases of matter or count degrees of freedom out-of equilibrium \cite{Srednicki:1993im,Kitaev:2005dm,Bhattacharya:2012mi}. Conformal field theories (CFT) are, on the other hand,  perfect testing grounds where one has analytic control over reduced density matrices what helps to understand the structure of entanglement at criticality \cite{Holzhey:1994we,Calabrese:2004eu}. In addition, via holography, CFTs at large central charge are dual to asymptotically Anti-de Sitter $(AdS)$ spacetimes \cite{Maldacena:1997re} where the celebrated Ryu-Takayangi formula \cite{RT} reproduces the von-Neumann entropy. 
  
\medskip  
Particularly useful measures of entanglement are provided by R\'enyi entanglement entropies (REE). For two regions $A$ and $B$, in state $\rho$ and a factorized Hilbert space $\mathcal{H}=\mathcal{H}_A\otimes \mathcal{H}_B$, one can define them for a reduced density matrix $\rho_A=\Tr_{\mathcal{H}_B}\rho$ as
\be
S^{(n)}_A=\frac{1}{1-n}\log\Tr_A\rho^n_A
\ee
where the integer power $n\ge 2$. From the family of all the REE we can extract the spectrum of the reduced density matrix but entropies for fixed $n$ are also interesting on their own, like e.g. purity for $n=2$ that might be turned into experimentally measurable quantities \cite{Islametal}. Finally, analytically continuing to non-integer $n$ and taking $n\to 1$ yields the celebrated von-Neumann entropy. 
 
\medskip
Based on the local quench protocol in CFTs \cite{cag,lquench}, it was demonstrated \cite{m2} that REEs are very useful in exploring entanglement also in excited states. More precisely, in a class of excited states obtained by acting with local operators on the CFT vacuum, the real time evolution of the change (with respect to the vacuum) in R\'enyi entanglement entropies $\Delta S^{(n)}_A$ can be used to characterize the operators from the perspective of quantum entanglement. An extensive study of $\Delta S^{(n)}_A$ in free field theories as well as minimal models shows that REEs approach to a constant that can be interpreted in terms of the quasi-particle picture \cite{m1, He:2014mwa}. On the other hand, in the limit of large central charge the entropies grow logarithmically with time in a universal fashion \cite{m3}. Some further results were obtained and extensively discussed in \cite{m4, LOP}.

\medskip
In quantum field theories, the presence of global conserved charges provides us with more parameters to tune when probing the system. It is then desirable to have appropriate quantum information measures sensitive to these charges in the entangled state. In fact, such charged R\'enyi entanglement entropy (CREE) was proposed in \cite{hchee}. It is defined as follows. If a theory has a global symmetry, a charge $Q$ can be defined. Subsequently, an entanglement charge localized in the subsystem $A$ is denoted as $Q_A$. On general grounds, a modular Hamiltonian $H_{mod}$ can be associated with the reduced density matrix as
\be
\rho_A = e^{-2\pi H_{mod}}
\ee 
Now, by taking the entanglement charge $Q_A$ into account, similarly to the grand canonical ensemble, a CREE is defined with respect to the charged reduced density matrix
\be
\rho_{A}^c=\f{e^{-2\pi H_{mod}+\alpha Q_A}}{\Tr_Ae^{-2\pi H_{mod}+\alpha Q_A}},
\ee
as the (integer) $n$-th R\'enyi entropy of $\rho_{A}^c$
\be
\begin{split}
S^{(n) ,c}_A =\f{1}{1-n}\log{\Tr_A \left(\rho^c_A\right)^n}.
\end{split}
\ee
Analogously to the (uncharged) R\'enyis, we can continue to the (charged) von-Neumann limit.

\medskip
So far, CREEs have only been studied in the CFT ground states \cite{dw1, mtu1} and we are only beginning to explore their properties and useful informations that they are sensitive to. To push this program further, in this work we will perform various computations, in CFTs in two and higher (even) dimensions as well as holography, on how CREEs evolve in time after local fermionic operator excitations. Before that, we first briefly review our setup and summarize the results below.

\subsubsection*{Our setup}
We consider excited states obtained by acting on the vacuum with local operators in free massless fermionic CFT in $D$ dimensional flat spacetime with Lorentzian signature. We choose half of the total space at $(t=0, x^1\ge0)$ as the subsystem $A$ and insert the local operator at distance $l$ from $A$ at time $t=-t$ (see Figure 1). The pure excited state is then defined as 
\begin{equation}\label{les}
\left|\Psi \right\rangle =\mathcal{N}\mathcal{O}(-t, - l, {\bf x})\left|0\right\rangle,
\end{equation}
where  $\mathcal{N}$ is the normalization constant such that the total density matrix $\rho = \ket{\Psi}\bra{\Psi}$ has a unit norm $\Tr (\rho)=1$ and ${\bf x}=(x_2, \cdots, x_{D-1})$ denotes the spatial coordinates perpendicular to the $x^1$. 
\begin{figure}[h!]
  \centering
  \includegraphics[width=8cm]{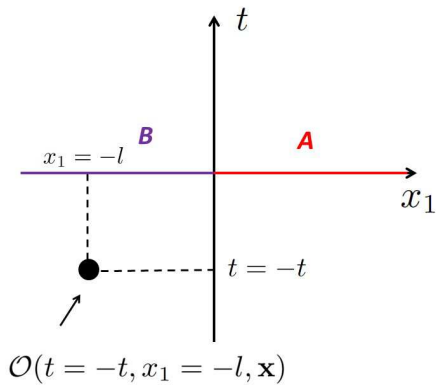}
  \caption{The location of local operator.}
\end{figure}
Tracing out the degrees of freedom in $B$ (the complement of $A$) we obtain the reduced density matrix $\rho_A = \Tr_B \rho$ of the sub-region $A$. \\
Using the replica trick, we can compute the change (with respect to the vacuum) in the R\'enyi as well as the charged R\'enyi entanglement entropies of $A$ due to the local operator excitation. They can be written in terms of the ratio of the $2n$-point correlator on $n$-sheeted surface $\Sigma_n$ with cuts corresponding to $A$ to the $n$-th power of the two-point function on a single sheet surface $\Sigma_1$  \cite{m4}
\begin{equation}
\begin{split}
\Delta S^{(n),c}_A =\frac{1}{1-n} \log{\left[\frac{\left \langle \mathcal{O}^{\dagger}(r, \theta^{n})\mathcal{O}(r_2, \theta_{2}^n)\cdots \mathcal{O}^{\dagger}(r,\theta^{1})\mathcal{O}(r_2,\theta_{2}^1) \right\rangle_{\Sigma_n}}{\left\langle\mathcal{O}^{\dagger}(r, \theta^1)\mathcal{O}(r_2,\theta_{2}^1) \right\rangle^n_{\Sigma_1}}\right]},
\end{split}
\end{equation}
where local operators are inserted periodically as in Figure 2 and $\theta^{k}=\theta+2(k-1)\pi$, $\theta_{2}^k=\theta_2+2(k-1)\pi$ and $k=1 \sim n$ $(k \in \mathbb{Z})$.
 \begin{figure}[h!]
  \centering
  \includegraphics[width=8cm]{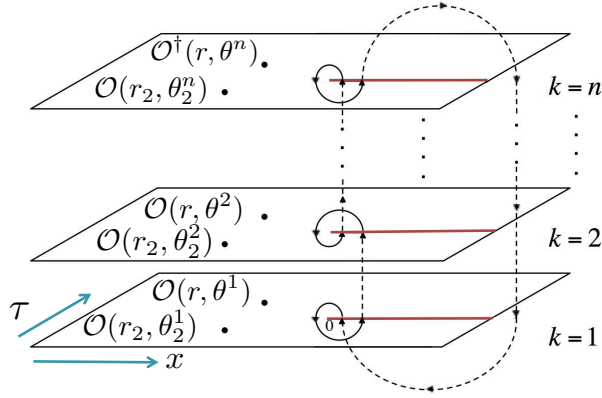}
  \caption{The n covering space $\Sigma_n$.}
\end{figure}
The difference between the standard R\'enyi entropies and the charged R\'enyi entropies is encoded in the action that computes the expectation values. Namely for charged R\'enyi entropies the action will be deformed by the appropriate charges associated with $A$. In our computations below we will then be interested in how $\Delta S^{(n),c}_A $ depends on the real time as well as the space-time dimension and the charge.

\subsubsection*{Summary}
Let us briefly summarize our computations and main results. Our first important result, is the determination of the twisted and untwisted fermionic propagators on $\Sigma_n$ in even dimensions. This allowed us to compute the excess in R\'enyi entanglement entropies in even dimensions and fix the dependence of the algebra of fermionic quasi-particles on the spacetime dimension (accomplishing the work started in \cite{m4}). 

\medskip
Moreover, using 2d CFT, the twisted propagators in four dimensions, the effective quasi particle description in arbitrary even dimensions as well as holography, we have managed to compute several examples of the change in CREEs. This important data gives us a lot of useful information about CREEs, how they evolve and depend on the charges. We were also able to interpret these results in terms of the charged fermionic quasi-particles that emerge as an effective, fine grained picture for local excitations as seen by CREEs.

\medskip
In this work, we especially focus on the dependence of CREEs on the modular chemical potential $\alpha$ in states excited by simple examples of fermionic e.g. $\bar{\psi} \psi$ and $\psi^{\dagger}\psi$. We found that for $\bar{\psi} \psi$ CREEs decrease when the modular chemical potential increases. On the other hand, for $\psi^{\dagger}\psi$ they initially increase at certain values of $\alpha$ and after they decrease. In both cases, increasing the potential, eventually brings CREEs to the value corresponding to a certain (dominant) charged sector. This property of CREEs is very similar to quantum information operations that through particular gates can access only a certain type of information from a quantum state. We believe that these features of CREEs may find important applications in the study of quantum entanglement in field theories.

\medskip
We have computed CREEs in free fermions in the adjoint representation of $U(N)$ and studied in detail the dependence on $N$, the modular chemical potential $\alpha$, the spacetime dimension $D$ and the replica number $n$. In the large $N$ limit entanglement in the "singlet sector" of the excited state seems to be enhanced. On the other hand, the large $\alpha$ CREEs have access to the internal "quark" sectors of the quantum state.
We have also noticed certain curious issues with the order of limits. In $D \neq 2$, after taking the large $N$ limit, the von Neumann limit ($n \rightarrow 1$) cannot be taken. On the other hand, in $D=2$ the limit $n \rightarrow 1$ can be taken after first taking the large modular chemical potential. 

\medskip
Last but not least, we studied the late time values of $\Delta S^{(n), c}_A$ holographically for $n\ge 2$. In dimensions higher than $3$ and in the approximation of the small modular chemical potential we have fund results consistent with \cite{m3} and the logarithmic growth with time but in a slightly modified way that depends on the charge. In addition, we found that in 3 dimensions with more analytic control, in a certain setup resembling the CREEs, the change in $\Delta S^{(n), c}_A$ can approach to a constant at late times.

\medskip
This paper is organized as follows. We begin section 2, by defining CREE and how to compute them using the replica trick. We find the relevant propagators that we employ in section 3 to compute REEs and determine the spacetime dependence of the quasi-particles that reproduce late time values of REEs in free fermionic theories. In the section 4, we finally compute and study the evolution and properties of $\Delta S^{(n), c}_A$ in various free fermionic field theories in even dimensions and holography in section 5. We finish with conclusions and discussion in section 6 and several relevant details and examples are included in four appendices.

\section{(Charged) R$\acute{e}$nyi entanglement entropies}
We start by explaining in detail the definition and how to compute charged (R$\acute{e}$nyi) entanglement entropy (CREE) in free fermions using replica trick.

\medskip
We consider a quantum field theory with global symmetries and corresponding conserved charges. In the computation of CREEs, we pick a spatial subsystem $A$ and its complement $B$ in a total pure state excited by the local operator. We ignore the degrees of freedom in B by tracing them out to obtain the reduced density matrix $\rho_A$. $Q_A$ is then defined as the (remaining) charge that is contained in the subsystem $A$. The charged reduced density matrix $\rho^c_A$ is by definition \cite{hchee}
\be
\begin{split} 
\rho^c_A = \f{e^{ \alpha Q_A}\rho_A }{\Tr_A e^{ \alpha Q_A}\rho_A },
\end{split}
\ee
where $\alpha$ is a modular chemical potential which in general can be real or purely imaginary. CREEs are defined for an integer $n\ge 2$ as REEs for the charged reduced density matrix
\be \label{cree}
\begin{split}
S^{(n) ,c}_A =\f{1}{1-n}\log{\Tr_A \left(\rho^c_A\right)^n}.
\end{split}
\ee
$S^{(n), c}_A$ can be computed using the replica trick generalizing \cite{Calabrese:2004eu}. In the path integral formalism, the effect of the modular chemical potential is captured by a Wilson line which encircles the entangling surface as in Figure 3.
\begin{figure}[h!]
  \centering
  \includegraphics[width=8cm]{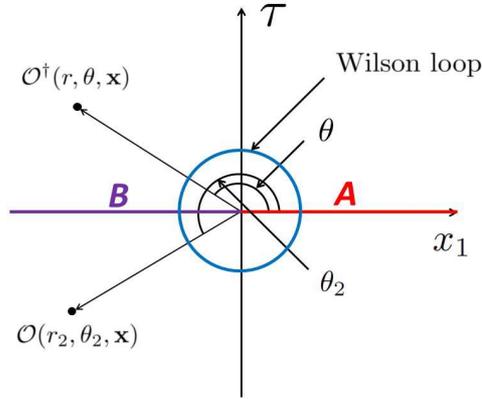}
  \caption{The location of local operator and Wilson line in the replica space.}
\end{figure}
The main quantity that we aim to compute is then the excesses of CREE of the subsystem $A$ due to the local fermionic operator
\be
\Delta S^{(n), c}_A = S^{(n), c}_{A, EX}- S^{(n), c}_{A, 0},
\ee
where $S^{(n), c}_{A, EX}$ and $S^{(n), c}_{A, 0}$ are CREEs of $A$ in the locally excited state and the ground state respectively.\\
As explained in the introduction, we study the real time dynamics of the excesses of CREE in the locally excited excited state. Our state is defined in flat space with Lorentzian signature and we follow the algorithm from the local quenches \cite{cag}. We add $e^{-i 2\pi \alpha_E Q_A}$ to the reduced density matrices where $\alpha_E$ is a real number. Namely, we first compute $\Delta S^{(n), c}_A$ by the replica trick in the Euclidean space.  Then, after computing the Green's function on the n-sheeted geometry $\Sigma_n$, we perform an analytic continuation to real time and change from pure imaginary chemical potential $-i \alpha_E$ to real chemical potential $\alpha$, as we will also explain in more detail later\footnote{The results for pure imaginary modular chemical potential are also presented in Appendix C}.\\

Clearly, if we take the chemical potentials to zero, all the above definitions and derivations for CREE reduce to the standard REE.

\subsection{The replica trick}    
We start with the Euclidean signature and add $e^{-2\pi i \alpha_E Q_A}$ to the reduced density matrix. Moreover, we assume that $0<\alpha_E\le 1$\footnote{Once we introduce a modular chemical potential $\alpha_E (\in {\bf R})$ to the reduced density matrix, it can be decomposed into the integer part $M \in \mathbb{Z}$ and fractional $F$ $(0<F\le 1)$. The boundary condition for fermions, that are sensitive to this chemical potential term, are only affected by the non-integer part $F$, therefore we neglect $M$.}. 
As a result, in the path integral, we introduce a Wilson line encircling entangling surface with a constant real gauge field from the imaginary chemical potential. Then, the action with which we compute CREEs with an imaginary entanglement chemical potential is given by
 \ba \label{aba}
S=\int dr r \int_{0}^{2n\pi}d\theta \int d{\bf x}^{D-2} \bar{\psi} \G^{\mu}  \nabla_{\mu} \psi +i\alpha_E \int dr r \int_{0}^{2n\pi}d\theta\int d{\bf x}^{D-2} \f {\bar{\psi}\G^{1}\psi}{r},
\ea
where $\G^{1} = i \g^t$ and the second term is just a convenient rewriting of the (fermionic) charge $Q_A$. In the following it is convenient to use the symmetry under the unitary redefinition as well as the local Lorentz transformation of the fermionic field. The fermion field can be rewritten as 
\begin{equation}\label{rw}
\psi(r, \theta, {\bf x})= e^{-i \alpha_E  \theta}\tilde{\psi}( r, \theta, {\bf x}),
\end{equation}
what brings the replica action to
\ba \label{ffa}
S=\int dr r \int_{0}^{2n\pi} d \theta \int d{\bf x}^{D-2} \bar{\tilde{\psi}} \G^{\mu}  \nabla_{\mu} \tilde{\psi}.
\ea
After performing the transformation  $\tilde{\psi} =e^{-\frac{\G^0\G^1}{2}\theta}\psi'$ where $\G^0=\g^1$, 
the action for $\psi'$ becomes
\ba
S=\int dr r \int_{0}^{2n\pi}d\theta \int d{\bf x}^{D-2} \bar{\psi}'\left[\G^r \rd _r+\G^{\theta}\frac{\rd_{\theta}}{r}+\G^{x}\rd_x\right] \psi'.
\ea
where $\Gamma$s in radial coordinates are given by
\begin{equation}
\begin{split}
&\G^{r}=\G^1 \sin{\theta}+\G^0\cos{\theta} \\
&\G^{\theta}=\G^1\cos{\theta}-\G^0\sin{\theta}.
\end{split}
\end{equation}
Finally, from this action, the Green's function for $\psi'$ can be written as a certain differential operator acting on a scalar function $G$
\begin{equation}
\begin{split}
S_{ab}(\theta-\theta_2)&=-\left\langle\mathcal{T} \psi'_a(r, \theta) \bar{\psi}'_b (r_2, \theta_2) \right \rangle \\
&=\left(\G^r\rd_r+\f{1}{r}\G^{\theta}\rd_{\theta}+\G^{\bf x}\rd_{\bf x}\right)_{ab}G(\theta-\theta_2).\label{gp}
\end{split}
\end{equation}
In addition, we impose a boundary condition for fermions as follows
\begin{equation}\label{bcf}
\begin{split}
\psi'(r,\theta+2 n \pi, {\bf x})= e^{i 2n\pi B}\psi'(r,\theta, {\bf x}), \\
\end{split}
\end{equation}
where $B$ is given by
\begin{equation}
B=\f{1-n(1-2\alpha_E)}{2n},
\end{equation}
that in the original fermion corresponds to
\be
 \psi(r, \theta+2n\pi)=-\psi (r, \theta).
\ee
\subsection{Propagators}
In this subsection, we compute the Green's function for $\psi'$ \cite{Line} in 4 dimensions. The function defined in (\ref{gp}) obeys the following equation of motion
\begin{equation}
\left(\G^r\rd_r+\frac{1}{r}\G^{\theta}\rd_{\theta}+\G^{\bf x}\rd_{\bf x}\right)S(\theta-\theta_2)=
- \f{\delta\left(\theta-\theta_2\right)\delta\left(r-r_2\right)\delta({\bf x}-{\bf x}_2 )}{r}{\bf 1},
\end{equation}
and consequently, $G(\theta-\theta_2)$ is a solution of
\begin{equation}
\left(\frac{\partial^2}{\partial r^2}+\frac{1}{r}\frac{\partial}{\partial r}+\frac{1}{r^2}\frac{\partial^2}{\partial \theta^2}+\frac{\partial^2}{\partial {\bf x}^2}\right)G(\theta-\theta_2)= - \f{\delta\left(\theta-\theta_2\right)\delta\left(r-r_2\right)\delta({\bf x}-{\bf x}_2 )}{r}.
\end{equation}
Both functions satisfy the same boundary conditions
\begin{equation}
S(r, \theta+2n \pi, {\bf x})=e^{i2n\pi B} S(r, \theta, {\bf x}),\qquad G(r, \theta+2n \pi, {\bf x})=e^{i2n\pi B} G(r, \theta, {\bf x}).
\end{equation}
We decompose $B$ into an integer and fractional parts
\begin{equation}
2n \pi B =2\pi (m + \alpha_r),
\end{equation} 
where $m\in \mathbb{Z}$ and $0< \alpha_r \le 1$\footnote{Here $a_r$ is given by $\alpha_r= \f{2n  \alpha_E - n +1 -2 m}{2}$} and 
the propagators only depend on $\alpha_r$.\\
Specifying to $D=4$, the propagator is given by 

\begin{equation}
G(x,x')=\frac{e^{i \frac{\alpha_r}{n}(\theta-\theta')}\sinh{\left(\f{1-\alpha_r}{n}t_0\right)}+ e^{i \frac{\alpha_r-1}{n}(\theta-\theta')}\sinh{\left(\f{\alpha_r}{n}t_0\right)}}{8n\pi^2rr'\sinh{\left(t_0\right)}\left(\cosh{\left(\f{t_0}{n}\right)}-\cos{\left(\f{\theta-\theta'}{n}\right)}\right)},
\end{equation}
where $t_0$ is defined by
\begin{equation}
\cosh{(t_0)} = \frac{r^2+r'^2+\left({\bf x}- {\bf x}'\right)^2}{2r r'}.
\end{equation}
The untwisted Green's functions ($\alpha_E=\f{N}{n}, N \in \mathbb{Z}, 0<N \le n$)\footnote{The boundary conditions for even or odd $n$ correspond to $\alpha_r=\f{1}{2} $ or $\alpha_r=0$ respectively.} reduce to those for REEs.\\
Then the propagators $S$ are simply obtained by applying the appropriate differential operator from the definition \eqref{gp}. 

\medskip
Finally, we perform an analytic continuation of the propagators in the Euclidean space to the Lorentzian signature. More precisely, the two local operators that appear in the reduced density matrix $\rho_A$ are located as in Figure 4. Therefore, the analytic continuation that yields the time dependent REEs and CREEs is given as follows \cite{m2}
\begin{equation}\label{ac}
\begin{split}
&\tau_l = \epsilon -i t, \\
&\tau_e=-\epsilon -i t,
\end{split}
\end{equation}
where in the Lorentzian signature $\epsilon$ is a cutoff parameter which regulates the divergence of the norm $\left\langle\Psi | \Psi\right\rangle$ of our locally excited state $\ket{\Psi}$.
\begin{figure}[h!]
  \centering
  \includegraphics[width=8cm]{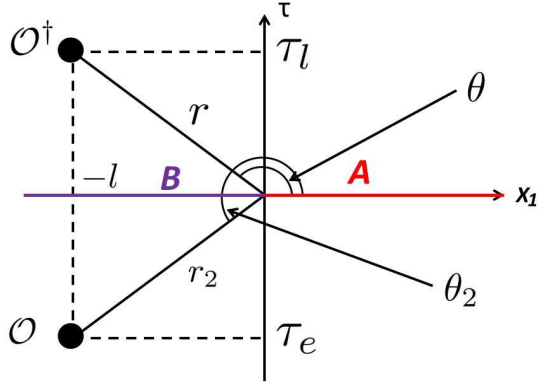}
  \caption{The relation between Lorentzian and Euclidean setup.}
\end{figure}
The relevant Lorentzian propagators can be found in Appendix A.

\medskip
In the next section, before computing charged Renyi entropies, we first use the untwisted propagator to analyze REEs. This will allow us to determine the dependence on the spacetime of the effective quasi-particle picture that we will later use to obtain CREEs in section 4.

\section{$\Delta S^{(n)}_A$ and dependence on spacetime dimension}
As explained earlier, authors in \cite{m4} pointed out that anti-commutation for entangled particles depends on the spacetime dimension.
In this section, we study it in detail. First we study $\Delta S^{(n)}_A$ for $\psi_a$,  $\bar{\psi}\psi$ and $\psi^{\dagger}\psi$ in $D=6, 8$ dim. spacetime by using propagators which are listed up in Appendix A\footnote{In even dimension, green functions can be computed analytically. In $D=2, 4$ $\Delta S^{(n)}_A$ have been already computed by the authors in \cite{m4}.   }.

\subsection{$\Delta S^{(n)}_A $ for $\psi_a$,  $\bar{\psi}\psi$ and $\psi^{\dagger}\psi$}
Here we study $\Delta S^{(n)}_A$ for $\psi_a$ , $\bar{\psi}\psi$ and $\psi^{\dagger}\psi$ in $6$ and $8$ dimensions by using the analytically continued propagators from Appendix A. After taking $\epsilon \rightarrow 0$, we also take the late time limit ($t\gg l$).  The computations are based on a straightforward application of Wick contractions so we only present the results for $\Delta S^{(n)}_A$ in Table 1. 

\begin{table}[h!]
  \begin{center}
\scalebox{0.8}{  
    \begin{tabular}{|c|c|c|} \hline
      $\mathcal{O}$ &   $\Delta S^{(n)}_A$ in 6 dim. & $\Delta S^{(n)}_A$ in 8 dim. \\ \hline \hline
      $\psi_a$ & $\f{1}{1-n}\log{\left[\left(\f{1}{4^n}\right)\left(2-\f{3}{4}\left(\gamma^1\gamma^t\right)_{aa}\right)^n+\left(\f{1}{4^n}\right)\left(2+\f{3}{4}\left(\gamma^1\gamma^t\right)_{aa}\right)^n \right]}$ &$ \f{1}{1-n} \log{\left[\left(\f{1}{2}+\f{5}{32}\left(\gamma^t\gamma^1\right)_{aa}\right)^n+\left(\f{1}{2}-\f{5}{32}\left(\gamma^t\gamma^1\right)_{aa}\right)^n\right]}$ \\ \hline
$\bar{\psi}\psi$  & $\f{1}{1-n}\log{\left[2\left(\f{1-\left(\f{3}{8}\right)^2}{4}\right)^n+2^3\left(\f{\left(\f{5}{4}\right)^2}{128}\right)^n+2^3\left(\f{\left(\f{11}{4}\right)^2}{128}\right)^n\right]}$ &$\f{1}{1-n}\log{\left[2\left(\f{1-\left(\f{5}{16}\right)^2}{4}\right)^n+2^4\left(\f{\left(\f{11}{16}\right)^2}{64}\right)^n+2^4\left(\f{\left(\f{21}{16}\right)^2}{64}\right)^n\right]}$ \\  \hline
      $\psi^{\dagger}\psi$ & $\f{1}{1-n}\log{\left[2\left( \f{1+\left(\f{3}{8}\right)^2}{4}\right)^n+2^4\left( \f{1-\left(\f{3}{8}\right)^2}{2^5}\right)^n\right]} $ & $\f{1}{1-n}\log{\left[2\left( \f{1+\left(\f{5}{16}\right)^2}{4}\right)^n+2^5\left( \f{1-\left(\f{5}{16}\right)^2}{2^6}\right)^n\right]}$ \\ \hline
	    \end{tabular}}
  \end{center}
   \caption{The list of the late time values of $\Delta S^{(n)}_A$ for $\ket{\Psi}=\mathcal{N}\mathcal{O}\ket{0}$ in $6$ and $8$ dimensions.}
\end{table}
Next we would like to interpret these results in terms effective density matrices for fermionic quasi-particles. We closely follow \cite{m4} where this was derived in 2 and 4 dimensions and we obtain density matrices listed in Table 2. 
\begin{table}[h!]
  \begin{center}
\scalebox{0.8}{  
    \begin{tabular}{|c|c|c|} \hline
      $\mathcal{O}$ &   $\rho_A$ in $6$ dim. & $\rho_A$ in $8$ dim. \\ \hline \hline
      $\psi_a$ & $\f{1}{2}\begin{pmatrix}
1-\f{3}{8}\left(\gamma^1\gamma^t\right)_{aa} & 0 \\
0 & 1+\f{3}{8}\left(\gamma^1\gamma^t\right)_{aa} \\
\end{pmatrix}$ &$ \f{1}{2}\begin{pmatrix}
1+\f{5}{16}\left(\gamma^t\gamma^1\right)_{aa} &0 \\
0 & 1-\f{5}{16}\left(\gamma^t\gamma^1\right)_{aa}
\end{pmatrix}$ \\ \hline
	      $\bar{\psi}\psi$  & $diag\left(\f{1-\left(\f{3}{8}\right)^2}{4},p_1, p_1, p_1, p_1, p_1, p_1, p_1, p_1, \f{1-\left(\f{3}{8}\right)^2}{4}\right)$ &$diag\left(\f{1-\left(\f{5}{16}\right)^2}{4},p_2, p_2, p_2, p_2, p_2, p_2, p_2, p_2, \f{1-\left(\f{5}{16}\right)^2}{4}\right)$ \\ 
 & where $p_1=diag\left(\f{5^2}{2^{11} }, \f{11^2}{2^{11}}\right)$ & where $p_2=diag\left(\f{11^2}{2^6 16^2}, \f{21^2}{2^6 16^2}, \f{11^2}{2^6 16^2}, \f{21^2}{2^6 16^2} \right)$$$ \\ \hline
    $\psi^{\dagger}\psi$  &  $diag\left(\f{1+\left(\f{3}{8}\right)^2}{4}, {\bf p}_3,  {\bf p}_3, \f{1+\left(\f{3}{8}\right)^2}{4}\right)$& $diag\left(\f{1+\left(\f{5}{16}\right)^2}{4}, {\bf p}_4,   {\bf p}_4,\f{1+\left(\f{5}{16}\right)^2}{4}\right)$ \\ 
&where $ {\bf p}_3=\f{1-\left(\f{3}{8}\right)^2}{2^5}{\bf 1}_{8\times 8}$ & where $ {\bf p}_4=\f{1-\left(\f{5}{16}\right)^2}{2^6}{\bf 1}_{16\times 16}$ \\ \hline
	    \end{tabular}}
  \end{center}
   \caption{The table of the reduced density matrix for $\ket{\Psi}=\mathcal{N}\mathcal{O}\ket{0}$ in $6$ and $8$ dimensions.}
\end{table}

Then $\Delta S^{(n)}_A$ in Table 1 can be reproduced from these effective density matrices by simply computing 
\be
\Delta S^{(n)}_A =\f{1}{1-n}\log{\rho_A^n},
\ee
In the next subsection, we will study how these density matrices and $\Delta S^{(n)}_A$ depend on the spacetime dimensions.

\subsection{The algebra of fermionic quasiparticles}
Here we interpret the results in the previous subsection in terms of quasi-particles (entanglement particles) (as in \cite{m4}) and fix its dependence on the spacetime dimensions.\\
We decompose fermionic fields into left movers ($\psi^{L\dagger}, \phi^{L\dagger}, \bar{\psi}^{L\dagger}$) and right movers ($\psi^{R\dagger}, \phi^{R\dagger}, \bar{\psi}^{R\dagger}$) as follows,
\begin{equation}\label{DQ}
\begin{split}
\psi_a&=\psi_a^{L\dagger}+\psi_a^{R\dagger}+\phi^{L}_a+\phi^{R}_a, \\
\psi_a^{\dagger}&=\psi_a^{L}+\psi_a^{R}+\phi^{L\dagger}_a+\phi^{R\dagger}_a, \\
\bar{\psi}_a&=\bar{\psi}^{L\dagger}_a+\bar{\psi}^{R\dagger}_a+i\left(\psi^L\gamma^t\right)_a+i\left(\psi^R\gamma^t\right)_a,
\end{split}
\end{equation}
where $\bar{\psi}^{K\dagger}$ is defined by
\begin{equation}
\begin{split}
\bar{\psi}^{L\dagger}_a \equiv i\left(\phi^{L\dagger}\gamma^t\right)_a,\\
\bar{\psi}^{R\dagger}_a \equiv i\left(\phi^{R\dagger}\gamma^t\right)_a.\\
\end{split}
\end{equation}
The vacuum is defined by
\be
\psi_a^{L}\ket{0}=\psi_a^{R}\ket{0}=\phi_a^{L}\ket{0}=\phi_a^{R}\ket{0}=\bar{\psi}_a^{L}\ket{0}=\bar{\psi}_a^{R}\ket{0}=0.
\ee 
The anti-commutation relation for them is given by
\begin{equation}\label{gdac}
\begin{split}
\{\phi^R_a, \phi^{R\dagger}_b\}&=\left(\delta_{ab}-c_g\left(\gamma^t\gamma^1\right)_{ab}\right), \\
\{\phi^L_a, \phi^{L\dagger}_b\}&=\left(\delta_{ab}+c_g\left(\gamma^t\gamma^1\right)_{ab}\right), \\
\{\psi^{R\dagger}_a, \psi^{R}_b\}&=\left(\delta_{ab}-c_g\left(\gamma^t\gamma^1\right)_{ab}\right), \\
\{\psi^{L\dagger}_a, \psi^{L}_b\}&=\left(\delta_{ab}+c_g\left(\gamma^t\gamma^1\right)_{ab}\right), \\
\{\bar{\psi}^R_a, \bar{\psi}^{R\dagger}_b\}&=\left(\delta_{ab}-c_g\left(\gamma^1\gamma^t\right)_{ab}\right), \\
\{\bar{\psi}^L_a, \bar{\psi}^{L\dagger}_b\}&=\left(\delta_{ab}+c_g\left(\gamma^1\gamma^t\right)_{ab}\right). \\
\end{split}
\end{equation}
where $c_g$ depends on the spacetime dimension. The $\g^t \g^1$ dependence in (\ref{gdac}) comes from the propagation of fermions and the fact that this influences the probability of (anti-)particles to be in $A$ and $B$ at late time. 
\begin{table}[htb]
  \begin{center}
    \begin{tabular}{|c|c|c|c|c|} \hline
      $D$ &   $2$& $4$& $6$& $8$ \\ \hline
      $c_g$ & $1$ & $\f{1}{2}$ & $\f{3}{8}$ & $\f{5}{16}$\\ \hline
	    \end{tabular}
	     \caption{The list of $c_g$ in $D$ dimensional spacetime.}
  \end{center}
\end{table}
To reproduce the result in the previous section, $c_g$ needs to be taken as in Table 3.
Here $D$ is the spacetime dimensions.
For general even spacetime dimension $D$, from the result in Table 3, $c_g$ is expected to be 
\be\label{dcog}
c_g=\f{1}{2^{D-2}}\sum_{j=0}^{\f{D-2}{2}}\left( {}_{\f{D-2}{2}}C_j\right)^2
\ee
where ${_l C_k}$ is the binomial coefficient\footnote{$c_g$ can be rewritten by $\f{1}{2^{D-2}}{_{D-2}C_{\f{D-2}{2}}}$ hence $c_g\le 1$.}.
\subsection{A physical interpretation of $c_g$}
Here we would like to interpret the coefficient ($c_g$) of $\gamma^t \gamma^1$.
It is defined in even dimensions and depends on $D$.  The probability of (anti-)particles' existence in $A$ or $B$ at late time is computable by using the anti-commutation in (\ref{gdac}). Therefore, the term
$c_g \g^t \g^1$ is related to the propagation of the (anti-)particles which are created by local operators. \\
In even dimension $D=2p~(p\in \mathbb{Z})$, fermions can be represented in the terms of $p$ spins' eigenvalues.
One of them can be chosen by that of $\gamma^t \gamma^1$. It determines how likely (anti-) particles are contained in $A$ or $B$ at the late time. On the other hand, the other eigenvalues do not affect this probability for (anti-)particles to be $A$ or $B$ at late times.
After fixing the direction of spin for $\gamma^t \gamma^1$, $\psi_a$ can be written by the linear combination of $\psi_a$ which are labeled by $q-1$ eigenvalues,  
\be
\begin{split}
\psi _a = N_1\psi_a (\gamma^t \gamma^1, \uparrow, \uparrow, \cdots, \uparrow) +N_2 \psi_a (\gamma^t \gamma^1, \uparrow, \uparrow, \cdots, \downarrow) + \cdots,
\end{split}
\ee
where $\uparrow$, $\downarrow$ represent eigenvalues of spins except of $\gamma^t \gamma^1$. 

The result in (\ref{dcog}) can be interpreted as follows.
The field which has $l$ negative eigenvalues can be written by $\Psi_a\left(\g^t \g^1, l, p-1-l \right)$.
There are ${_{p-1-l}C_l} $ of $\Psi_a\left(\g^t \g^1, l, p-1-l \right)$.
$\psi_a$ is approximately given by the sum of them such as
\be \label{comf}
\begin{split}
\psi_a \sim \sum_{l=0}^{p-1}{_{p-1-l}C_l} \Psi_a\left(\g^t \g^1, l, p-1-l \right).
\end{split}
\ee
The field which has the same number of negative values can contribute to the probability of  (anti-)particles in $A$ or $B$ at the late time equally.
Then the binomial coefficient included in $c_g$ is expected to come from the number of $\Psi_a\left(\g^t \g^1, l, p-1-l \right)$. 
Therefore square of binomial coefficient appears in $c_g$. 
As explained above, the coefficient of $\g^t\g^1$ is expected to depend on how to divide the total system into $A$ and $B$.
If another shape is taken as the subsystem $A$ (for example, $x_1\ge 0, x_2\ge 0$), the dependences on other spins (for example $\g^t\g^2$) are expected to  appear in $c_g$. Here we assume that the volume of the subsystem $A$ is infinite.

\section{The excess of charged (R$\acute{e}$nyi) entanglement entropy}
In this section we study $\Delta S^{(n), c}_A$ for a locally excited state in free massless fermionic field theories. As explained above, their late time values can be reproduced from a quantum state in an effective Hilbert space for quasi-particles which are created by local operators. Those entanglement entropies on the effective Hilbert apace are finite. Therefore it is expected that studying $\Delta S^{(n,) c}_A$ for the locally excited state helps us to understand the properties of CREEs as well as the charge carried by the quasi-particles. 

First we study $\Delta S^{(n), c}_A$ in $2$d massless free fermion in 2d CFT and interpret the results in terms of charged quasi-particles. Next, we consider $\Delta S^{(n), c}_A$ in $4$d free massless fermionic field theory using twisted propagators and also the effective quasi-particle picture. Finally, we generalize the computation to free fermions with $U(N)$ symmetry in even $D$ dimensions.

\subsection{2d CFT: massless free fermions}

We start by considering a Dirac fermion (or two Majorana fermions ) $\psi = (\varphi, \ti{\bar{\varphi}})^T$ (and the Dirac conjugate $\bar{\psi} = (\ti{\varphi}, \bar{\varphi})$), where $\varphi$ and $\bar{\varphi}$ ( $\ti{\varphi}$ and $\ti{\bar{\varphi}}$  ) is the combination of Majorana-Weyl fermion $\psi_1(z) $ and $\psi_2(z)$ ($\ti{\psi}_1(\bar{z})$ and $\ti{\psi}_2(\bar{z})$)
\ba
\varphi = \psi_1 + i \psi_2, \  \ \  \ \ti{\varphi} = \ti{\psi}_1 + i \ti{\psi} _2 \no
\bar{\varphi}= \psi_1 - i \psi_2 ,\ \ \ \ \ti{\bar{\varphi}}=\ti{\psi}_1 - i\ti{\psi}_2.
\ea
This system has a $U(1)$ global symmetry $\varphi \to e^{i \theta} \varphi$ ($\ti{\varphi} \to e^{ -i \theta} \ti{\varphi}$) and we choose the charge for this symmetry as $Q_A$. In the following we employ the bosonization
\ba
\varphi (z) \simeq e^{iH(z) } , \ \ \ \ \ti{\varphi }(\bar{z}) \simeq e^{ i \ti{H}(z)} ,\no
\bar{\varphi}(z) \simeq e^{- iH(z)}, \ \ \ \ \ti{\bar{\varphi}} (\bar{z}) \simeq e^{- i \ti{H}(\bar{z})},
\ea
where each of $H(z)$ and $H(\bar{z})$ is the chiral and anti-chiral part of $\phi(z,\bar{z})$.

Now, let us compute the second CREE for the case of local operator
 $\mcl{O}=\frac{1}{\sqrt{2}}( e^{i \phi} + e^{- i \phi})$. In the fermion language, this operator corresponds the Lorentz invariant operator $\bar{\psi}\psi$ .\\
The correlation function on $\Sigma_2$ is given by
\ba
\f{\left\langle\mcl{O}^{\dagger}(w_1)\mcl{O}(w_2)\mcl{O}^{\dagger}(w_3) \mcl{O}(w_4)e^{- i \alpha \phi}(w_5) e^{i \alpha \phi}(w_6)\right\rangle_{\Sigma_2}}{\left\langle e^{- i \alpha \phi}(w_5) e^{i \alpha \phi}(w_6)\right\rangle_{\Sigma_2}} .
\ea
Here $w_5 = 0$ and $w_6 = L$ are the branch points and finally we take the limit of $w_6 \to \infty$ because we are interested in the case that $A$ is a half line. 
Then we map the 2-sheeted surface $\Sigma_2$ to a complex plane $\Sigma_1$ by the conformal transformation $z^2 = w/(w-L)$. 
The operator $e^{\pm i \alpha \phi(w_i)}$ represents the existence of flux.
We need to treat the twist operator $e^{ \pm i \alpha \phi}$ located at the branch point carefully. First, these operators are inserted at singular points, the conformal factor $|dw/dz| $ diverges. But these factors are cancelled by conformal factor from normalization part. Second, we need to treat the chemical potential $\alpha$ carefully.
Because we introduced these operators to represent the flux corresponding to the chemical potential, they also need to reproduce the correct monodromy. \\
If we go around the twist operator once in $\Sigma _1$ , we go around twice in $\Sigma_2$, 
and the operator gets the monodromy transformation $e^{\pm i  H(z)} \to e^{\pm 2 \alpha i} e^{\pm i H(z)}$.
To reproduce this condition, we need to change the flux $\alpha $ to $2 \alpha$. 
Taking these into account, the correlation function on $\Sigma_2 $ can be expressed 
as follows:
\ba
&&\f{\left\langle\mcl{O}^{\dagger}(w_1)\mcl{O}(w_2)\mcl{O}^{\dagger}(w_3) \mcl{O}(w_4)e^{- i \alpha \phi}(w_5) e^{i \alpha \phi}(w_6)\right\rangle_{\Sigma_2}}{\left\langle e^{- i \alpha \phi}(w_5) e^{i \alpha \phi}(w_6)\right\rangle_{\Sigma_2}}  \notag \\
&&= \prod_{i= 1}^{4} \Big| \f{d w_i}{dz_i} \Big | ^{- 2 \Delta_i} \f{\left\langle\mcl{O}^{\dagger}(z_1)\mcl{O}(z_2)\mcl{O}^{\dagger}(z_3) \mcl{O}(z_4)e^{- 2 i \alpha \phi}(z_5) e^{2i \alpha \phi}(z_6)\right\rangle_{\Sigma_1}}{\left \langle e^{-2 i   \alpha \phi}(z_5) e^{ 2i \alpha \phi}(z_6)\right\rangle_{\Sigma_1}} \\ \notag
\ea
Using Appendix \ref{App:2d}, we can compute the ratio of propagators
\be
\begin{split}
&\f{\left\langle\mcl{O}^{\dagger}(w_1) \mcl{O}(w_2)\mcl{O}^{\dagger}(w_3) \mcl{O}(w_4)\tilde{\mu} (w_5) \mu(w_6) \right\rangle_{\Sigma_2}/\left \langle \tilde{\mu}(w_5) \mu(w_6)\right\rangle_{\Sigma_2}}{
(\left\langle\mcl{O}^{\dagger}(w_1) \mcl{O}(w_2) \tilde{\mu}(w_5)\mu(w_6)\right\rangle_{\Sigma_1}/
\left\langle \tilde{\mu}(w_5)\mu(w_6\right\rangle_{\Sigma_1})^2
} \\
&= \f{|\f{z_1}{z_2}| ^{8 \alpha} + |\f{z_2}{z_1}|^{8 \alpha} + 2|\f{z_1 + z_2}{2}|^8 + 2|\f{z_{12}}{2}|^8
}{
(|\f{z_1}{z_2}| ^{4 \alpha} + |\f{z_2}{z_1}|^{4 \alpha})^2
}
\end{split}
\ee
with $\mu(w_i)=e^{i \alpha \phi}(w_i)$ and $\tilde{\mu}(w_i) = e^{- i \alpha \phi(w_i)}$.\\
Then, if we take the limit of $\ep \to 0$ , we can get the early and late time values of CREE. However, we need to take care of the complex coordinates $z$ and $\bar{z}$ separately since, after the analytical continuation to real time (or in the Lorentzian regime), they are not just complex conjugates of each other. The coordinates of operators (after the mapping $w = f(z)$) $z_1 , \cdots , z_4$ are 
\be
z_1 = - z_3 = \s{\f{l - t - i\ep}{ l + L - t - i \ep}} ,\ \ \ \bar{z}_1 = - \bar{z}_3 = \s{\f{l + t - i\ep}{ l + L + t - i \ep}}
\ee
\be
z_2 = - z_4 = \s{\f{l - t + i\ep}{ l + L - t + i \ep}},\ \ \ \bar{z}_1 = - \bar{z}_3 = \s{\f{l + t - i\ep}{ l + L + t - i \ep}}.
\ee
We take the branch as follows
\ba
z _1 = \lim _{\ep \to 0} \s{\f{l -t - i\ep}{l + L - t - i\ep}}  = \s{\f{l -t }{l + L - t }},  \ \ \  (0 < t < l) \\
z _2 = \lim _{\ep \to 0} \s{\f{l -t + i\ep}{l + L - t + i\ep}}  = \s{\f{l -t }{l + L - t }}  \ \ \  (0 < t < l),
\ea
and then , if $l < t < L$ there should be a phase factor
\ba
z _1 = \lim _{\ep \to 0} \s{\f{l -t - i\ep}{l + L - t - i\ep}}  = e ^{-\f{i \pi}{2}}\s{\f{t -l }{l + L - t }},  \ \ \  (0 < t < l) \\
z _2 = \lim _{\ep \to 0} \s{\f{l -t + i\ep}{l + L - t + i\ep}}  = e^{ \f{i\pi}{2}} \s{\f{t - l }{l + L - t }}  \ \ \  (0 < t < l),
\ea
so we need to take these phase factors into account (there are no phase factor for $\bar{z}_1$). Namely, in the range of $l < t < L$, 
\be
\lim _{\ep \to 0}\Big|\f{z _1}{z _2} \Big| ^{2\alpha}  = e^{- i \alpha \pi}.
\ee
Finally, from these results, the excess of 2nd charged R\'enyi entropy is given by 
\ba
\Delta S^{(2),c}_A(\alpha) &=& -\log{\left[
\f{\left\langle\mcl{O}^{\dagger}(w_1) \mcl{O}(w_2)\mcl{O}^{\dagger}(w_3) \mcl{O}(w_4)\sigma (w_5) \sigma(w_6) \right\rangle_{\Sigma_2}/\left\langle\sigma(w_5) \sigma(z_6)\right\rangle_{\Sigma_2}}{
(\left\langle\mcl{O}^{\dagger}(w_1) \mcl{O}(w_2) \sigma(w_5)\sigma(w_6)\right\rangle_{\Sigma_1}/
\left\langle\sigma(w_5)\sigma(w_6\right\rangle_{\Sigma_1})^2
}\right]} \notag \\ 
&=&
\begin{cases}
0 & (0 < t < l) \\
- \log\left( \f{\cos 4 \pi \alpha}{2 (\cos 2 \pi \alpha) ^2}\right) & (l < t < L)
\end{cases}.
\ea
In the next subsection we provide a physical interpretation of this result in terms of charged quasi-particles.

\subsubsection{Particle interpretation}
In \cite{m2}, it was shown that (R\'enyi) entanglement entropy for excited states by local operators can be understood as 
the entropy of quasi-particles created by the local operator.  We can also see that the above result can be understood as CREE of a
charged quasi-particles created by the operator $\mcl{O}$. In the case of $\mcl{O} = 1/\s{2}(e^{i\phi} + e^{-i\phi})$, we define a state 
\be
\mcl{O}\left|0 \right\rangle \equiv \frac{1}{\sqrt{2}} (\left| q_A\right\rangle\left| q_{\bar{A}}\right\rangle + \left| -q_A\right\rangle\left| -q_{\bar{A}}\right\rangle)
\ee
such that the reduced density matrix becomes  
\be
\rho _A = \f{1}{2} (\left| q_A\right\rangle \left\langle q_A\right| + \left| -q_A\right\rangle \left\langle -q_A\right|) .
\ee
Now, we define the charge operator as 
\ba
&&Q = Q_A \otimes Q_{\bar{A}} \\  &&Q _A \left|q_A\right\rangle = q\left| q_A\right\rangle \\ &&Q_A \left| -q_A\right\rangle = - q\left| -q_A\right\rangle
\ea
Then, the charged R\'enyi entropy becomes 

\be
S^{(n),c}_A(\alpha) = \f{1 }{1 -n} \log \f{\tr(\rho_A e^{i \alpha Q_A})^n}{ (\tr(\rho_A e^{i \alpha Q_A}))^n} = \f{1}{1-n} \log \f{\cos  n \alpha q }{ 2 ^{n-1} (\cos  \alpha q ) ^n }. \label{imaginarycp}
\ee\\
If we substitute $q = 2\pi$ and $n=2$, the value of charged R\'enyi entropy in the quasi-particle state reproduces the answer from the replica computation in 2d CFT.

\begin{figure}[h!]
  \centering
  \includegraphics[width=9cm]{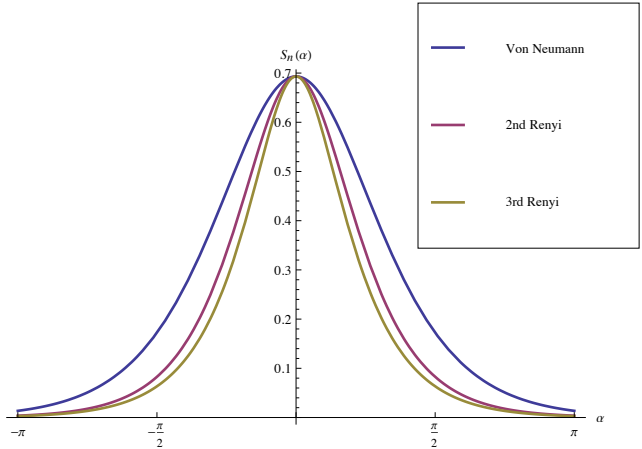} 
  \caption{The plots of $S^{(n),c}_A(\alpha)$ for $n=1,2,3$ as functions of entanglement chemical potential $\alpha$. } 
\label{fig:plot2dCREE}
\end{figure}

Moreover, we can also get the charged R\'enyi entropy for real chemical potential
\be
S^{(n),c}_A(\alpha) = \f{1 }{1 -n} \log \f{\tr(\rho_A e^{ \alpha Q_A})^n}{ (\tr(\rho_A e^{\alpha Q_A}))^n} = \f{1}{1-n} \log \f{\cosh  n \alpha q }{ 2 ^{n-1} (\cosh  \alpha q ) ^n }.
\ee 
This can be obtained by the analytic continuation from (\ref{imaginarycp}) around $\alpha = 0$. As functions of entanglement chemical potential, charged entanglement entropy and charged R\'enyi entropy behave as in Figure \ref{fig:plot2dCREE}. 
 \\
The important point is that in $|\alpha| \to \infty$ limit CREEs vanish. This is because for the real entanglement chemical potential, 
only one of the charged sectors: $\ket{q_A}$ or  $\ket{-q_A}$, essentially contributes and the state is seen by CREEs as effectively not entangled.

\subsection{Free fermions in $4$ dim.}
In this subsection, we study the time evolution for the excesses of $\Delta S^{(n), c}_A$ in $4$d massless free fermionic field theory using twisted propagators explained in Appendix A. We add real modular chemical potential to the reduced density matrix. For that, we first we compute $\Delta S^{(n), c}_A$ with imaginary modular chemical potential by using analytically continued propagators (from Appendix A) and the limit $\epsilon \rightarrow 0$. Then we perform the analytic continuation in the modular chemical potential  $(\alpha_E \rightarrow i \alpha)$, to obtain $\Delta S^{(n), c}_A $ with real modular chemical potential.

\medskip
To illustrate the properties of $\Delta S^{(n), c}_A$, we consider states excited by three simple fermionic operators: a component of Dirac fermion $\psi_a$, a Lorentz scalar $\bar{\psi}\psi$ (where $\bar{\psi} = i\psi^{\dagger} \g^t $) and $\psi^{\dagger}\psi$. As we will see, their properties under Lorentz transformations also play an important role from the perspective of (charged) quantum entanglement.

In the region  $0<t<l$, $\Delta S^{(n), c}_A$ vanish for all of the operators. This is consistent with the propagation of the quasi particles that by that time are still in the traced-out region. On the other hand, for times $t \ge l$, $\Delta S^{(n), c}_A$ shows the initial growth with time and finally saturates to a constant values for the late time ($t \gg l$).  The initial growths with time, as well as the late time constants are different for each operator and we list them in Table 4 and Table 5 respectively.
\begin{table}[htb]
  \begin{center}
    \begin{tabular}{|c|c|} \hline
      $\mathcal{O}$ &   $\Delta S^{(n),c}_A$ for $t\ge l$  \\ \hline \hline
      $\psi_a$ & $\f{1}{1-n}\log{\left[\f{(t+l)^n\left[2-\frac{l-t}{t}\left(\g^t \g^1\right)_{aa}\right]^n+(t-l)^n\left[2-\frac{l+t}{t}\left(\g^t \g^1\right)_{aa}\right]^n e^{-2n \pi \alpha}}{\left[2\left\{\left(t-l\right)e^{-2\pi \alpha}+(t+l)\right\}-\frac{(-l+t)(l+t)}{t}\left(e^{-2\pi \alpha}-1\right)\left(\g^t \g^1\right)_{aa}\right]^n}\right]}$ \\ \hline
      $\bar{\psi}\psi$  & $\f{1}{1-n}\log{\left[  \f{(t+l)^{2n}\left[4^2-4\left(\f{t-l}{t}\right)^2\right]^n+(t-l)^{2n}\left[4^2-4\left(\f{t+l}{t}\right)^2\right]^n+8(t^2-l^2)^n\sum^{n\ge k}_{k \in 2{\bf Z}}{}_n C_k4^k\left(4+\f{t^2-l^2}{t^2}\right)^{n-k}\cosh{\left(2n\pi \alpha\right)}}{\left[4^2\left\{(t-l)^2+(t+l)^2+2(t^2-l^2)\cosh{\left(2\pi \alpha\right)}\right\}+8\left(\f{t^2-l^2}{t}\right)^2(\cosh{\left(2\pi \alpha\right)}-1)\right]^n}\right]} $  \\  \hline
      $\psi^{\dagger}\psi$ & $\f{1}{1-n}\log{\left[\f{(t+l)^{2n}\left[4^2+4\left(\f{t-l}{t}\right)^2\right]^n+(t-l)^{2n}\left[4^2+4\left(\f{t+l}{t}\right)^2\right]^n+8(t^2-l^2)^n\sum^{n\ge k}_{k \in 2{\bf Z}}{}_n C_k\left(4\f{l}{t}\right)^k\left(4+\f{-t^2+l^2}{t^2}\right)^{n-k}\cosh{\left(2n\pi \alpha\right)}}{\left[4^2\left\{(t-l)^2+(t+l)^2+2(t^2-l^2)\cosh{\left(2\pi \alpha\right)}\right\}+8\left(\f{t^2-l^2}{t}\right)^2(-\cosh{(2\pi \alpha)}+1)\right]^n}\right]} $  \\ \hline
    \end{tabular}
     \caption{The list of $\Delta S^{(n)}_A$ for $\ket{\Psi}=\mathcal{N}\mathcal{O}\ket{0}$ in the region of initial growth $t\ge l$.}
  \end{center}
\end{table}

\begin{table}[htb]
  \begin{center}
    \begin{tabular}{|c|c|} \hline
      $\mathcal{O}$ &   $\Delta S^{(n),c}_A$ for $t\gg l$  \\ \hline \hline
      $\psi_a$ & $\f{1}{1-n}\log{\left[\f{\left[2+\left(\g^t\g^1\right)_{aa}\right]^n+\left[2-\left(\g^t\g^1\right)_{aa}\right]^ne^{-2n \pi \alpha}}{\left[2\left(e^{- 2 \pi \alpha}+1\right)-\left(e^{- 2 \pi \alpha}-1\right)\left(\g^t\g^1\right)_{aa}\right]^n}\right]}$\\ \hline
      $\bar{\psi}\psi$  & $\f{1}{1-n}\log{\left[ \frac{(12)^n+(12)^n+8 \sum_{k \in 2{\bf Z}}^{n \ge k}{}_n C_k5^{n-k}4^k\cosh{\left(2n\pi \alpha\right)}}{\left[4^2\left(2+2\cosh{(2\pi \alpha)}\right)+8\left(\cosh{\left(2\pi \alpha\right)}-1\right)\right]^n}\right]}$  \\  \hline
      $\psi^{\dagger}\psi$ & $\f{1}{1-n}\log{\left[ \frac{(20)^n+(20)^n+8\cdot 3^n\cosh{\left(2n\pi \alpha\right)}}{\left[4^2\left(2+2\cosh{(2\pi \alpha)}\right)+8\left(-\cosh{\left(2\pi \alpha\right)}+1\right)\right]^n}\right]}$  \\ \hline
    \end{tabular}
    \caption{The table of $\Delta S^{(n)}_A$ for $\ket{\Psi}=\mathcal{N}\mathcal{O}\ket{0}$ at late times $t\gg l$.}
  \end{center}
\end{table}
Next, we interpret the time evolution and the late time values of $\Delta S^{(n), c}_A$ in these tables in terms of charged quasi-particles.

\subsubsection{Particle interpretation}
Here we interpret the evolution of $\Delta S^{(n),c}_A$ from the previous section in terms of quasi-particles.
We decompose $\psi_a$, $\psi_a^{\dagger}$ and $\bar{\psi}_a$ into right and left movers respectively as in (\ref{DQ}) and define states in the effective Hilbert space by acting with such decomposed operators. At the late time, we assume that the degrees of freedom of the left and right movers are identified with the degrees of freedom in $B$ and $A$ respectively. In $4$d, anti-commutation relations for the left and the right movers are given by (\ref{gdac}) with $c_g = \f{1}{2}$. 
Then, the reduced density matrix is defined by
\be\label{rdm2}
\begin{split}
\rho_A=\f{\Tr_B\left( \mathcal{O}\ket{0}\bra{0}\mathcal{O}^{\dagger}\right)}{\Tr\left( \mathcal{O}\ket{0}\bra{0}\mathcal{O}^{\dagger}\right)}, ~~~~
\end{split}
\ee
where $\Tr_A \rho_A=1$. 
The reduced density matrices $\rho_A^p, \rho_A^{bp}, \rho^{dp}_A $ for $\psi_a, \bar{\psi}\psi , \psi^{\dagger} \psi$ are then computed as in \cite{m4} and are listed in Table 6.
\begin{table}[h!]
  \begin{center}
    \begin{tabular}{|c|c|} \hline
      $\mathcal{O}$ &   $\rho_A$  \\ \hline \hline
      $\psi_a$ & $\frac{1}{4}\left[\left(2+\left(\gamma^t\gamma^1\right)_{aa}\right)\left|0\right\rangle_R\left\langle0\right|_R+\left(2-\left(\gamma^t\gamma^1\right)_{aa}\right)\left|\psi^R_a \right\rangle_R\left\langle\psi^R_a \right|_R \right]$\\ \hline
      $\bar{\psi}\psi$  & $\frac{12}{64}\left|0\right\rangle_R\left\langle0\right|_R+\sum_{i}\frac{\mathcal{M}_i}{64}\left|\psi^R_i\right\rangle_R\left\langle \psi^R_i\right|_R+\sum_{i}\frac{\bar{\mathcal{M}}_i}{64}\left|\bar{\psi}^R_i\right\rangle_R\left\langle\bar{ \psi}^R_i\right|_R+\frac{12}{64}\left|\bar{\psi}^R\psi^R\right\rangle_R\left\langle\bar{\psi}^R\psi^R \right|_R$  \\  \hline
      $\psi^{\dagger}\psi$ & $\f{20}{64}\left| \phi^R \psi^R \right \rangle_R \left \langle \phi^R \psi^R \right|_R +\f{20}{64}\left| 0 \right \rangle_R \left \langle 0 \right|_R
+\f{3}{64}\sum_i \left|\phi^R_i\right\rangle_R\left\langle \phi^R_i \right|_R+\f{3}{60}\left|\psi^R_i\right\rangle_R\left\langle \psi^R_i \right|_R$  \\ \hline \hline
 &   $\rho_A^c$  \\ \hline \hline
      $\psi_a$ & $C_{0, R}\ket{0}_R\bra{0}_{R}+C_{1, R}\ket{\psi^R_a}_R\bra{\psi^R_a}_R,$\\ \hline
      $\bar{\psi}\psi$  & $D_{0, R}\ket{0}_R\bra{0}_{R}+\sum_a D_{1, R,  a}\ket{\psi^R_a}_R\bra{\psi^R_a}_R+\sum_a D_{2, R, a}\ket{\bar{\psi}^R_a}_R\bra{\bar{\psi}^R_a}_R+D_{3, R}\left|\bar{\psi}^R\psi^R\right\rangle_R\left\langle\bar{\psi}^R\psi^R \right|_R$  \\  \hline
      $\psi^{\dagger}\psi$ & $E_{0, R}\left| \phi^R \psi^R \right \rangle_R \left \langle \phi^R \psi^R \right|_R +E_{1, R}\left| 0 \right \rangle_R \left \langle 0 \right|_R
+E_{2, R}\sum_i \left|\phi^R_i\right\rangle_R\left\langle \phi^R_i \right|_R++E_{3, R}\sum_i \left|\psi^R_i\right\rangle_R\left\langle \psi^R_i \right|_R$  \\ \hline
    \end{tabular}
  \end{center}
  \caption{The table of $\rho_A$ and $\rho^c_A$ for $\ket{\Psi}=\mathcal{N}\mathcal{O}\ket{0}$ in the late time. Here $\mathcal{M}_1=\mathcal{M}_3=\bar{\mathcal{M}}_2=\bar{\mathcal{M}}_4=1$ and $\mathcal{M}_2=\mathcal{M}_4=\bar{\mathcal{M}}_1=\bar{\mathcal{M}}_3=9$. $\rho_A$ was obtain by authors in \cite{m4}. The probabilities $C, D$ and $E$ are listed up in Table 7.}
\end{table}

\begin{table}[h!]
  \begin{center}
    \begin{tabular}{|c|c|c|c|} \hline
      $C_{0, R}$ & $\f{\left(2+\left(\g^t\g^1\right)_{aa}\right)}{\left[2(1+e^{-2\pi \alpha})+(1-e^{-2\pi \alpha})\left(\g^t\g^1\right)_{aa}\right]}$ &$C_{1, R}$ &$\f{e^{-2 \pi \alpha}\left(2-\left(\g^t\g^1\right)_{aa}\right)}{\left[2(1+e^{-2\pi \alpha})+(1-e^{-2\pi \alpha})\left(\g^t\g^1\right)_{aa}\right]}$ \\ \hline \hline
   $D_{0, R}=D_{3, R}$ & $\f{12}{24+40 \cosh{\left(2\pi \alpha\right)}}$ &$D_{1, R, 1}=D_{1, R, 3}$ &$\f{9 e^{-2\pi \alpha}}{24+40 \cosh{\left(2\pi \alpha\right)}}$ \\ \hline
$D_{1, R, 2}=D_{1, R, 4}$ & $\f{ e^{-2\pi \alpha}}{24+40 \cosh{\left(2\pi \alpha\right)}}$ &$D_{2, R, 1}=D_{2, R, 3}$& $ \f{ e^{2\pi \alpha}}{24+40 \cosh{\left(2\pi \alpha\right)}}$ \\ \hline 
$D_{2, R, 2}=D_{2, R, 4}$ & $\f{9 e^{2\pi \alpha}}{24+40 \cosh{\left(2\pi \alpha\right)}}$ & & \\ \hline \hline
$E_{0, R}=E_{1, R}$ & $\f{20}{40+24\cosh{(2\pi \alpha)}}$ &$E_{2, R}$ &$\f{3e^{-2 \pi \alpha}}{40+24\cosh{(2\pi \alpha)}}$ \\ \hline
$E_{3, R}$ & $\f{3e^{2 \pi \alpha}}{40+24\cosh{(2\pi \alpha)}}$ & & \\ \hline
    \end{tabular}
  \end{center}
  \caption{The table of $C, D$ and $E$.}
\end{table}
In the free massless fermionic field theory, we have a $U(1)$ global charge. Therefore an entanglement charge $Q_A$ can be defined by the charge included in the region $A$. The charged reduced density matrix $\rho_A^c$ is defined by 
\be
\begin{split}
\rho_A^c = \f{e^{2\pi \alpha Q_A}\rho_A}{\Tr_A e^{2\pi \alpha Q_A}\rho_A}.
\end{split}
\ee
where $\rho_A$ is defined in (\ref{rdm2}) and $\Tr_A\rho_A^c=1$.\\
The dependence of operators on $Q_A$ are given by
\be
\begin{split}
[Q_A, \psi_a^{R \dagger}]=- \psi_a^{R \dagger},~~~~[Q_A, \bar{\psi}^{R \dagger}_a]=\bar{\psi}^{R \dagger}_a,~~~~[Q_A, \phi^{R \dagger}_a]=\phi^{R \dagger}_a
\end{split}
\ee 
where we assume that the subsystem $A$ is $x^1\ge 0$. We also assume that $\g^t\g^1=diag(1, -1, 1, -1)$ for simplicity. The reduced density matrices $\rho_A^{p, c} $, $\rho_A^{bp, c}$ and $\rho_A^{dp, c}$ are listed in the Table 6. If we compute $\Delta S^{(n), c}_A$ by using the reduced density matrices in the Table 6, they are consistent with those in the Table 5.

\medskip
If we take the von Neumann limit $n \rightarrow 1$, $\Delta S^{c}_A$ for $\psi_a, \bar{\psi}\psi$ and $\psi^{\dagger} \psi$ are respectively given by
\be
\begin{split}
&\Delta S^{p, c}_A =\f{2+\left(\g^t\g^1\right)_{aa}}{F_1}\log{\left[\f{F_1}{2+\left(\g^t\g^1\right)_{aa}}\right]}+\f{\left(2-\left(\g^t\g^1\right)_{aa}\right)e^{-2\pi\alpha}}{F_1}\log{\left[\f{F_1}{\left(2-\left(\g^t\g^1\right)_{aa}\right)e^{-2\pi\alpha}}\right]}, \\
&\Delta S^{bp, c}_A =\f{3}{F_2}\log{\left[\f{2F_2}{3}\right]}+\f{5\cosh{\left(2\pi\alpha\right)}}{F_2}\log{\left[8F_2\right]}-\f{9\cosh{2\pi\alpha}}{F_2}\log{3}-\f{10\pi\alpha\sinh{2\pi\alpha}}{F_2}, \\
&\Delta S^{dp, c}_A=\f{5}{F_3}\log{\left[\f{2F_3}{5}\right]}+\f{3\cosh{\left(2\pi\alpha\right)}}{F_3}\log{\left[\f{8F_3}{3}\right]}-\f{6\pi\alpha\sinh{2\pi\alpha}}{F_3}, \\
\end{split}
\ee
where the functions $F_i$ are given by
\be
\begin{split}
&F_1=2+\left(\g^t\g^1\right)_{aa}+e^{-2\pi\alpha}(2-\left(\g^t\g^1\right)_{aa}),\\
&F_2=3+5\cosh{2\pi\alpha}, \\
&F_3=5+3\cosh{2\pi\alpha}.
\end{split}
\ee
$\Delta S^{p, c}_A, \Delta S^{bp, c}_A$ and $\Delta S^{dp, c}_A$ respectively correspond to the excesses of charged entanglement entropies for $\psi_a$, $\bar{\psi}\psi$ and $\psi^{\dagger}\psi$. 

Let us next study in detail how the late time values of $\Delta S^{(n), c}_A$ depend on the modular chemical potential.  

\subsubsection*{The dependence on modular chemical potential}
Here we study how the late time values of $\Delta S^{(n), c}_A$ for $\psi_a, \bar{\psi} \psi$ and $\psi^{\dagger} \psi$ depend on the modular chemical potential $\alpha$.
$\Delta S^{c}_A$ are plotted in Figures 6 and 7. The Figure 7 show that $\Delta S^{c}_A$ for $\bar{\psi}\psi$ monotonically decreases when $\left|\alpha\right|$ increases. 
\begin{figure}[h!]
  \centering
  \includegraphics[width=8cm]{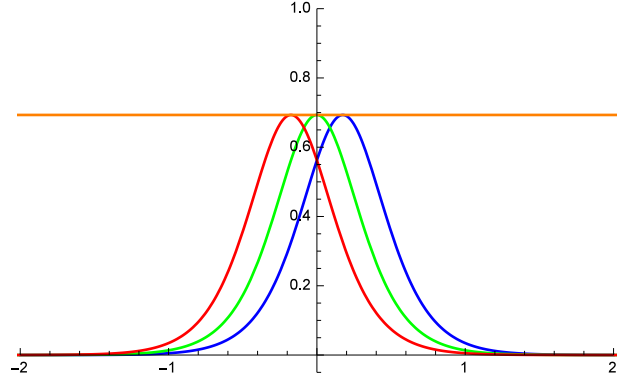}
  \caption{The plot of $\Delta S^{c}_A$ for $\psi_a$ with real modular chemical potential $\alpha$.The horizontal axis corresponds to $\alpha$.The vertical line corresponds to $\Delta S^{c}_A$.  $\psi_a$  acts on the ground state. The blue, green and red lines respectively correspond to $\left(\g^t \g^1\right)_{aa}=-1, 0$ and $1$. The orange line corresponds to $\log 2$.}
\end{figure}

\begin{figure}[h!]
  \centering
  \includegraphics[width=8cm]{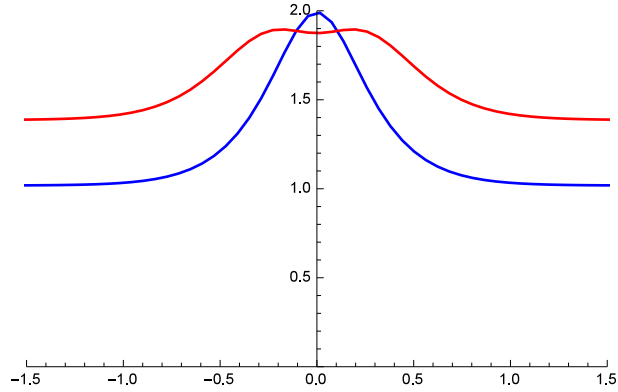}
  \caption{The plot of local $\Delta S^{c}_A$ for $\bar{\psi}\psi$ and $\psi^{\dagger}\psi$ with real modular chemical potential $\alpha$.The horizontal axis corresponds to the modular chemical potential $\alpha$. The vertical line corresponds to $\Delta S^{c}_A$. $\psi^{\dagger}\psi$ acts on the ground state. The blue and red curves correspond to $\Delta S^{c}_A$ for $\bar{\psi}\psi$ and $\psi^{\dagger}\psi$ respectively}
\end{figure}
On the other hand, Figure 6 and 7 show that  $\Delta S^{c}_A$ for $\psi_a~(\left(\g^t\g^1\right)_{aa}\neq 0), \psi^{\dagger} \psi$ do not monotonically decrease when $\left|\alpha\right|$ increases.
In both case, until certain values $\left|\alpha\right|$ they increase and decrease after that value. 
$\Delta S^{c}_A$ for $\psi^{\dagger}\psi$ reaches maximum value at $\alpha_m$ which satisfies the following equation\footnote{However it is difficult to compute the maximal value of $\Delta S^{c}_A$ for $\psi^{\dagger}\psi$ analytically.},
\be
2 \pi  \alpha_m  (5 \cosh (2 \pi  \alpha_m )+3)-5 \log \left(\frac{20}{3}\right) \sinh (2 \pi  \alpha_m )=0.
\ee
In the large $\left|\alpha \right|$ limit, $\Delta S^{(n), c}_A$ for $\psi_a, \bar{\psi}\psi$ and $\psi^{\dagger}\psi$ eventually approach a certain value. They are listed in Table 8.
These results show that the dependence of $\Delta S^{c}_A$ on modular chemical potential is related the detail of local operators which act on the ground state. 
For a certain class of local operators, the amount of quantum information can increase by tuning modular chemical potential. It is expected that operators can be characterized by measuring $\Delta S^{(n), c}_A$. It is one of the interesting future problems.

Next, we consider the behavior of $\Delta S^{(n), c}_A$ in the large $|\alpha|$ limit.

\subsubsection*{A limit of large modular chemical potential}
Here, we focus on the behavior of $\Delta S^{(n), c}_A$ in the large $\left| \alpha \right|$ limit, which, as we will see, can be used to extract entanglement in a certain charged sector.

If we consider large modular chemical potential ($\alpha \rightarrow \pm \infty$), we can easily extract CREEs e.g. $\Delta S^{(n), c}_A$, $\Delta S^{c}_A$ and $\Delta S^{(\infty), c}_A$  for $\psi_a$, $\bar{\psi}\psi$ and $\psi^{\dagger}\psi$ and they are listed in the Table 8. 
\begin{table}[htb]
  \begin{center}
    \begin{tabular}{|c|c|c|c|} \hline
      $\mathcal{O}$ &  $\Delta S^{(n), c}_A,$ & $\Delta S^{c}_A$ & $\Delta S^{(\infty), c}_A$ \\ \hline \hline
      $\psi_a$ &  $0$ & $0$ & $0$ \\ \hline
      $\bar{\psi}\psi$ &  $\frac{\log \left(\frac{2\ 9^n+2}{20^n}\right)}{1-n}$ & $\log \left(\frac{20}{3\ 3^{4/5}}\right)$ & $\log{\frac{20}{9}}$ \\ \hline
     $\psi^{\dagger}\psi$ &  $\log{4}$ & $ \log{4}$ & $\log{4}$ \\ \hline
    \end{tabular}
  \end{center}
  \caption{The table of $\Delta S^{(n), c}_A, \Delta S^{c}_A$ and $\Delta S^{(\infty), c}_A$ for $\ket{\Psi}=\mathcal{N}\mathcal{O}\ket{0}$ when adding the large modular chemical potential.}
\end{table}
Let us explain this behavior for each of the operators.\\
First, if we take the  $\alpha \rightarrow \pm \infty$ limit for operator $\psi_a$, $\Delta S^{(n), c}_A$ reduce to that for product state $\psi_a^{R \dagger} \ket{0}_R$ (or $\ket{0}_R$ ) and vanishes. \\
If $\bar{\psi}\psi$ acts on the ground state, in the large modular chemical potential limit, the reduced density matrix becomes 
\be
\begin{split}
\rho^{pb, c}_A \rightarrow \begin{cases}
\sum_a D_{2, R, a}\ket{\bar{\psi}^R_a}_R\bra{\bar{\psi}^R_a}_R & \alpha \rightarrow \infty\\
\sum_a D_{1, R,  a}\ket{\psi^R_a}_R\bra{\psi^R_a}_R & \alpha \rightarrow -\infty, \\
\end{cases}
\end{split}
\ee
where
\be
\begin{split}
&D_{1, R, 1}=D_{1, R, 3}=\f{9}{20},~~
D_{1, R, 2}=D_{1, R, 4}=\f{1}{20},~~
D_{2, R, 1}=D_{2, R, 3}=\f{1}{20}, ~~
D_{2, R, 2}=D_{2, R, 4}=\f{9 }{20}. \\
\end{split}
\ee
Therefore $\Delta S^{c}_A$ corresponds to the answer in a given charge sector and therefore in the large $\alpha$ limit, entanglement in a particular charged sector can be extracted. \\
Finally, in the large $\left|\alpha \right|$ limit, the reduced density matrix for $\psi^{\dagger}\psi$ is given by
\be
\begin{split}
\rho^{pb, c}_A \rightarrow \begin{cases}
E_{2, R}\sum_i \left|\phi^R_i\right\rangle_R\left\langle \phi^R_i \right|_R & \alpha \rightarrow \infty\\
E_{1, R}\sum_i \left|\psi^R_i\right\rangle_R\left\langle \psi^R_i \right|_R & \alpha \rightarrow -\infty\\
\end{cases}
\end{split}
\ee
where
\be
\begin{split}
&E_{1, R}=\f{1}{4},~~ E_{2, R}=\f{1}{4}. \\
\end{split}
\ee
Therefore $\Delta S^{c}_A$ for the large $\alpha$ is given by that for the maximally entangled state in a certain charged sector. 
This way, we can extract quantum information in a particular charged sector by turning the large modular chemical potential $\alpha$. In the large modular chemical potential limit  $e^{2\pi\alpha Q_A}$ acts on the ground state as projection operator which extracts the effect of some sectors.\\
It is also worth noting that without the modular chemical potential, $\Delta S^{ c}_A$ for $\bar{\psi}\psi$ is greater than that for  $\psi^{\dagger}\psi$. On the other hand in the large $\alpha$ limit, $\Delta S^{c}_A$ for $\psi^{\dagger}\psi$ overwhelms that for $\bar{\psi}\psi$. This is another new feature of the CREEs that was not present in REEs.

\subsection{$D$-dimensional free-fermions in the adjoint of $U(N)$}

In this subsection, we generalize the setup to the $D$ dimensional free massless fermion with $U(N)$ symmetry in the Lorentzian signature,
\be
S=-\int d^Dx\, \Tr\left[ \bar{\psi}(x)\g^{\mu}\partial_{\mu} \psi(x)\right],
\ee
where $\psi$ is the adjoint fermion. 
Here we consider $\Delta S^{(n)}_A$ for $(\psi_a)^l_k, \Tr\left(\bar{\psi}\psi\right)$ and $\Tr\left(\psi ^{\dagger} \psi\right)$.
In $D$ dim. $U(N)$ free massless fermionic field theory, fermionic fields are decomposed into the left and right moving modes as in (\ref{DQ}) and anti-commutation relation for them are imposed as follows,

\begin{equation}\label{gdac2}
\begin{split}
\left\{\left(\phi^R_a\right)^l_k, \left(\phi^{R\dagger}_b\right)^n_m\right\}&=\delta^l_m\delta^n_k\left(\delta_{ab}-c_g\left(\gamma^t\gamma^1\right)_{ab}\right), \\
\left\{\left(\phi^L_a\right)^l_k, \left(\phi^{L\dagger}_b\right)^n_m\right\}&=\delta^l_m\delta^n_k\left(\delta_{ab}+c_g\left(\gamma^t\gamma^1\right)_{ab}\right), \\
\left\{\left(\psi^{R\dagger}_a\right)^l_k, \left(\psi^{R}_b\right)^n_m\right\}&=\delta^l_m\delta^n_k\left(\delta_{ab}-c_g\left(\gamma^t\gamma^1\right)_{ab}\right), \\
\left\{\left(\psi^{L\dagger}_a\right)^l_k, \left(\psi^{L}_b\right)^n_m\right\}&=\delta^l_m\delta^n_k\left(\delta_{ab}+c_g\left(\gamma^t\gamma^1\right)_{ab}\right), \\
\left\{\left(\bar{\psi}^R_a\right)^l_k, \left(\bar{\psi}^{R\dagger}_b\right)^n_m\right\}&=\delta^l_m\delta^n_k\left(\delta_{ab}-c_g\left(\gamma^1\gamma^t\right)_{ab}\right), \\
\left\{\left(\bar{\psi}^L_a\right)^l_k, \left(\bar{\psi}^{L\dagger}_b\right)^n_m\right\}&=\delta^l_m\delta^n_k\left(\delta_{ab}+c_g\left(\gamma^1\gamma^t\right)_{ab}\right), \\
\end{split}
\end{equation}
where $c_g$ is defined by (\ref{dcog}). 

Let's derive $\Delta S^{(n)}_A$ and $\Delta S^{(n), c}_A$ for $\left(\psi_a\right)^i_j\ket{0}, \Tr\bar{\psi}\psi\ket{0}$ and $\Tr\psi^{\dagger}\psi \ket{0}$ as in the previous subsection.

\subsubsection{$\Delta S^{(n)}_A$ and $\Delta S^{(n), c}_A$ for $\left(\psi_a\right)^i_j$, $\Tr(\bar{\psi}\psi)$ and $\Tr(\psi^{\dagger}\psi)$}
\begin{table}[htb]
\begin{center} 
\scalebox{0.9}{  
 \begin{tabular}{|c|c|c|}\hline
      $\mathcal{O}$ & $\Delta S^{(n \ge2)}_A$ for $D\neq 2$ &$\Delta S^{(n \ge2)}_A$ for $D=2$ \\ \hline
      $\left(\psi_a\right)^{i}_j$ & $\f{1}{1-n}\log{\left[\f{\left(1+c_g \left(\g^t \g^1\right)_{aa}\right)^n+\left(1- \left(\g^t \g^1\right)_{aa}\right)^n}{2^n}\right]}$ & $\f{1}{1-n}\log{\left[\f{\left(1+ \left(\g^t \g^1\right)_{aa}\right)^n+\left(1- \left(\g^t \g^1\right)_{aa}\right)^n}{2^n}\right]}$ \\ \hline
      $\Tr\bar{\psi}\psi$  & $\f{1}{1-n}\log{\left[\f{2\cdot2^{\f{D n}{2}}(1-c_g^2)^n N^{2n}+2^{\f{D}{2}}(1-c_g)^{2n}N^{2}+2^{\f{D}{2}}(1+c_g)^{2n}N^{2}}{N^{2n}2^{\f{(D+4)n}{2}}}\right]}$ & $\log{2N^2}$\\ \hline
     $\Tr\psi^{\dagger}\psi$   &$\f{1}{1-n}\log{\left[\f{2\cdot 2^{\f{D n}{2}}\left(1+c_g^2\right)^n N^{2n}+2\cdot 2^{\f{D}{2}}(1-c_g^2)^n N^2}{2^{\left(\f{D+4}{2}\right)n}N^{2n}}\right]}$& $\log{2}$ \\ \hline \hline
 $\mathcal{O}$ &  $\Delta S^{(n \ge2), c}_A$ for $D\neq2$&$\Delta S^{(n \ge2), c}_A$ for $D=2$  \\ \hline
$\left(\psi_a\right)^{i}_j$ & $\f{1}{1-n}\log{\left[\f{\left(1+c_g \left(\g^t \g^1\right)_{aa}\right)^n+e^{- 2 \pi \alpha n}\left(1-c_g \left(\g^t \g^1\right)_{aa}\right)^n}{\left(1+e^{-2\pi \alpha}+c_g (1- e^{-2 \pi \alpha})\left(\g^t \g^1\right)_{aa}\right)^n}\right]}$ & $\f{1}{1-n}\log{\left[\f{\left(1+ \left(\g^t \g^1\right)_{aa}\right)^n+e^{- 2 \pi \alpha n}\left(1- \left(\g^t \g^1\right)_{aa}\right)^n}{\left(1+e^{-2\pi \alpha}+ (1- e^{-2 \pi \alpha})\left(\g^t \g^1\right)_{aa}\right)^n}\right]}$\\ \hline
$\Tr\bar{\psi}\psi$  &$\f{1}{1-n}\log{\left[\f{2\cdot 2^{\f{D n}{2}}(1-c_g^2)^nN^{2n}+2^{\f{D}{2}}N^2\cosh{\left(2n \pi \alpha\right)\left\{(1-c_g)^{2n}+(1+c_g)^{2n}\right\}}}{2^{\f{(D+2)n}{2}}N^{2n}\left\{\left(1-c_g^2\right)+(1+c_g^2)\cosh{\left(2\pi \alpha\right)}\right\}^n}\right]}$ & $\log{2N^2}+\f{1}{1-n}\log{\left[\f{\cosh{2n\pi \alpha}}{\left(\cosh{2\pi\alpha}\right)^n}\right]}$ \\ \hline
$\Tr\psi^{\dagger}\psi$   &$\frac{1}{1-n}\log \left[\frac{2 \left(c_g^2+1\right)^n 2^{\frac{D n}{2}} N^{2 n}+2^{\frac{D+2}{2}} N^2 \left(1-c_g^2\right)^n \cosh (2 \pi  \alpha  n)}{2^{\frac{1}{2} (D+2) n} N^{2 n} \left(\left(1-c_g^2\right) \cosh (2 \pi  \alpha )+\left(c_g^2+1\right)\right)^n}\right]$ &$\log{2}$ \\ \hline
    \end{tabular}}
\end{center}
 \caption{The table of $\Delta S^{(n\ge2)}_A$ and $\Delta S^{(n\ge2), c}_A$ for $\left(\psi_a\right)^i_j\ket{0}, \Tr\bar{\psi}\psi\ket{0}$ and $\psi^{\dagger}\psi \ket{0}$.}
\end{table}

\begin{table}[htb]
\begin{center}   
\scalebox{0.7}{  
 \begin{tabular}{|c|c|c|} \hline
      $\mathcal{O}$ & $\Delta S_A$  for $D\neq 2$  & $\Delta S_A$ for $D =2$  \\ \hline
      $\left(\psi_a\right)^{i}_j$ &$\f{1+c_g\left(\g^t\g^1\right)_{aa}}{2}\log{\left(\f{2}{1+c_g\left(\g^t\g^1\right)_{aa}}\right)}+\f{1-c_g\left(\g^t\g^1\right)_{aa}}{2}\log{\left(\f{2}{1-c_g\left(\g^t\g^1\right)_{aa}}\right)}$ & $\f{1+\left(\g^t\g^1\right)_{aa}}{2}\log{\left(\f{2}{1+\left(\g^t\g^1\right)_{aa}}\right)}+\f{1-\left(\g^t\g^1\right)_{aa}}{2}\log{\left(\f{2}{1-\left(\g^t\g^1\right)_{aa}}\right)}$ \\ \hline
      $\Tr\bar{\psi}\psi$   & $\f{(1-c_g^2)}{2}\log{\left(\f{4}{1-c_g^2}\right)}+\f{(1-c_g)^2}{4}\log{\left(\f{2^{\f{D+4}{2}}N^2}{(1-c_g)^2}\right)}+\f{(1+c_g)^2}{4}\log{\left(\f{2^{\f{D+4}{2}}N^2}{(1+c_g)^2}\right)}$ &$\log{2N^2}$ \\ \hline
$\Tr\psi^{\dagger}\psi$   & $\f{1+c_g^2}{2}\log{\left[\f{4}{1+c_g^2}\right]}+\f{(1-c_g^2)}{2}\log{\left[\f{N^22^{\f{D+2}{2}}}{1-c_g^2}\right]}$ & $\log{2}$  \\ \hline \hline
 $\mathcal{O}$ & $\Delta S^{c}_A$ for $D \neq2$ & $\Delta S^{c}_A$ for $D =2$   \\ \hline
$\left(\psi_a\right)^{i}_j$ & $\f{1+c_g\left(\g^t\g^1\right)_{aa}}{F_3}\log{\left(\f{F_3}{1+c_g\left(\g^t\g^1\right)_{aa}}\right)}+\f{\left(1-c_g\left(\g^t\g^1\right)_{aa}\right)}{F_3e^{2\pi \alpha}}\log{\left(\f{F_3e^{2\pi \alpha}}{\left(1-c_g\left(\g^t\g^1\right)_{aa}\right)}\right)}$ & $\f{1+\left(\g^t\g^1\right)_{aa}}{F_4}\log{\left(\f{F_4}{1+\left(\g^t\g^1\right)_{aa}}\right)}+\f{\left(1-\left(\g^t\g^1\right)_{aa}\right)}{F_4e^{2\pi \alpha}}\log{\left(\f{F_4e^{2\pi \alpha}}{\left(1-\left(\g^t\g^1\right)_{aa}\right)}\right)}$\\
 & where $F_3=1+c_g\left(\g^t\g^1\right)_{aa}+(1-c_g\left(\g^t\g^1\right)_{aa})e^{-2\pi \alpha}$ & $F_4=1+\left(\g^t\g^1\right)_{aa}+(1-\left(\g^t\g^1\right)_{aa})e^{-2\pi \alpha}$\\ \hline
$\Tr\bar{\psi}\psi$  & $\f{(1-c_g^2)}{F_1}\log{\left(\f{2F_1}{(1-c_g^2)}\right)}+\f{(1-c_g)^2\cosh{(2\pi \alpha)}}{2F_1}\log{\left(\f{2^{\f{D+2}{2}}N^2F_1}{(1-c_g)^2}\right)}$ & $$ \\ 
& $+\f{(1+c_g)^2\cosh{(2\pi \alpha)}}{2F_1}\log{\left(\f{2^{\f{D+2}{2}}N^2F_1}{(1+c_g)^2}\right)}-\f{2\pi \alpha (1+c_g^2)\sinh{2\pi \alpha}}{F_1}$ & $\log{\left(2N^2\cosh{2\pi \alpha}\right)}-2\pi \alpha \tanh{2\pi \alpha}$ \\
& where $F_1=(1-c_g^2)+(1+c_g^2)\cosh{2\pi \alpha}$ & $$ \\ \hline
$\Tr\psi^{\dagger}\psi$   & $\f{1+c_g^2}{F_2}\log{\left(\f{2F_2}{1+c_g^2}\right)}+\f{(1-c_g^2)\cosh{(2\pi \alpha)}}{F_2}\log{\left( \f{N^22^{\f{D+2}{2}}F_2}{(1-c_g^2)}\right)}$ & $$  \\
   &~~~~~~~~~~~~~~~~~~~~~~~~~~~~~~$-2\pi \alpha \f{(1-c_g^2)\sinh{2\pi\alpha}}{F_2}$  & $\log{2}$ \\
 & where $F_2=1+c_g^2+(1-c_g^2)\cosh{2\pi \alpha}$ & $$ \\ \hline
    \end{tabular}}
\end{center}
\caption{The table of $\Delta S^c_A$  for $\left(\psi_a\right)^i_j\ket{0}, \Tr\bar{\psi}\psi\ket{0}$ and $\Tr\psi^{\dagger}\psi \ket{0}$.}
\end{table}

Here we study $\Delta S^{(n)}_A$ and $\Delta S^{(n), c}_A$ for  $\left(\psi_a\right)^i_j\ket{0}, \Tr\bar{\psi}\psi\ket{0}$ and $\Tr\psi^{\dagger}\psi \ket{0}$ in even $D$ dim. spacetime. Those for $n \ge 2$ are listed up in the Table 9\footnote{When $D=2$ and $\left(\g^t\g^1\right)_{aa}=\pm 1$, $\Delta S^{(n)}_A$ and $\Delta S^{(n), c}_A$ for $\left(\psi_a\right)^i_j$ vanish.}.

 $\Delta S_A$ and $\Delta S^{c}_A$ for $\left(\psi_a\right)^i_j\ket{0}, \Tr\bar{\psi}\psi\ket{0}$ and $\Tr\psi^{\dagger}\psi \ket{0}$ are listed up in the Table 10.

The results in the Table 9 and 10 show that $\Delta S^{(n)}_A$ for $\left(\psi_a\right)^i_j$ do not depend on $N$.  They show that $\Delta S^{(n)}_A$ for $\Tr \bar{\psi} \psi$ do not depend on the replica number $n$ for $D=2$. $\Delta S^{(n), c}_A$ for $\Tr \psi ^{\dagger}\psi$ do not depend on $\alpha$ and $n$ for $D=2$. We would like to explain why $\Delta S^{(n)}_A$ for $\Tr \bar{\psi} \psi $ ($\Delta S^{(n)}_A$ for $\Tr \psi^{\dagger} \psi $) do not depend on $n$ ($n$ and $\alpha$) in $D=2$ in the next subsection.
In this subsection let's study $\Delta S^{(n)}_A$ and $\Delta S^{(n), c}_A$ in the large $N$ or $\alpha$ limit. First let's study them in the large $N$ limit. After that, let's study them in the large $\alpha$ limit. 

\subsubsection*{Large $N$ Limit}

Here we consider $\Delta S^{(n)}_A$ for $\Tr \bar{\psi}\psi$ and $\Tr \psi^{\dagger} \psi$ in the large $N$ limit. When $D \neq 2$, $\Delta S^{(n)}_A$ for $\Tr \bar{\psi}\psi$ in the large $N$ limit 
 are given by
\be
\begin{split}\label{nbp}
&\Delta S^{(n \ge 2)}_A\sim\f{1}{1-n}\log{\left[2^{1-2n}(1-c_g^2)^n\right]}, \\
&\Delta S_A \sim \left(c_g^2+1\right) \log {(N)}.
\end{split}
\ee
On the other hand, $\Delta S^{(n)}_A$ for $\Tr \psi^{\dagger}\psi$ in the large $N$ limit are
given by 
\be
\begin{split}\label{ndp}
&\Delta S^{(n \ge 2)}_A\sim\f{1}{1-n}\log{\left[2^{1-2n}(1+c_g^2)^n\right]}, \\
&\Delta S_A \sim \left(-c_g^2+1\right) \log {(N)},
\end{split}
\ee
where $\Delta S^{(n \ge 2)}_A$ are consistent with the results in \cite{m4} for $D=4$.
In the large $N$ limit, $\Delta S_A$ for $\Tr \psi^{\dagger}\psi$ is smaller than $\Delta S_A$ for $\Tr \psi^{\dagger}\psi$.
In both case, the coefficient of $\log{N}$ depends on $c_g$. Then it depends on spacetime dimension $D$.

For $D=2$, $\Delta S^{(n)}_A$ for $\Tr \bar{\psi} \psi$ is given by 
\be
\Delta S^{(n)}_A \sim \log{N}.
\ee
As in Table 9 and 10, $\Delta S^{(n)}_A$ for $\Tr\psi^{\dagger}\psi$ do not depend on $N$ and are given by 
\be \label{uiN}
\Delta S^{(n)}_A =\log{2}.
\ee
For $\Tr \bar{\psi} \psi$ and $\Tr \psi^{\dagger} \psi$ in $D>2$, von Neumann limit $(n \rightarrow 1)$ can not be taken after taking the large $N$ limit.
However in $D=2$ for $\Tr \bar{\psi} \psi$ and $\Tr \psi^{\dagger} \psi$ von Neumann limit $(n \rightarrow 1)$ can be taken after taking the large $N$ limit. As explain later, in $D=2$ case there is only one sector where components of the reduced density matrix equally depend on $N$. Then $n\rightarrow 1$ limit can be taken after taking $N \rightarrow \infty$.

We would also like to study $\Delta S^{(n), c}_A$ in the large $N$ limit.
First let's consider $\Delta S^{(n), c}_A$ for $\Tr\bar{\psi}\psi$.
In the limit $N \rightarrow \infty$ for $D \neq 2$, 
they reduce to 
\be
\begin{split}\label{rst1}
&\Delta S^{(n\ge2), c}_A \sim \f{1}{1-n}\log{\left[\f{2(1-c_g^2)^n}{2^n\left\{(1-c_g^2)+(1+c_g^2)\cosh{\left(2\pi \alpha\right)}\right\}^n}\right]}, \\
&\Delta S^{c}_A\sim \frac{2\cosh (2 \pi  \alpha ) \left(c_g^2+1\right) }{ \left(\left(c_g^2+1\right) \cosh (2 \pi  \alpha )-c_g^2+1\right)}  \log (N)\\
\end{split}
\ee
The results in (\ref{rst1}) show that after taking $N \rightarrow \infty$, von Neumann limit can not be taken for $\Delta S^{(n\ge 2), c}_A$ in $D(>2)$ dimensional spacetime.

If we take the large $N$ limit for $D=2$, $\Delta S^{(n), c}_A$ for arbitrary $n$ reduce to
\be\label{D2Nc}
\Delta S^{(n), c}_A \sim 2\log{N}.
\ee
The result in (\ref{D2Nc}) indicates $\Delta S^{(n), c}_A$ for $D=2$ in the large $N$ limit do not depend on the modular chemical potential. On the other hand, the coefficient of $\log{N}$ in $D \neq 2$ depends on the modular chemical potential $\alpha$. In the next subsection, we would like to explain how the coefficient of $\log{N}$ depends on $\alpha$.
  
In the limit $N \rightarrow \infty$, $\Delta S^{(n), c}_A$ for $\Tr\psi^{\dagger}\psi$ in $D \neq 2$ reduce to 
\be
\begin{split}
&\Delta S^{(n \ge 2), c}_A \sim\f{1}{1-n}\log{\left[\f{2(1+c_g^2)^n}{\left[2\left\{(1+c_g^2)+(1-c_g^2)\cosh{\left(2\pi \alpha\right)}\right\}\right]^n}\right]}, \\
&\Delta S^{c}_A\sim\frac{2\left(c_g^2-1\right) \cosh (2 \pi  \alpha ) }{ \left(c_g^2-1\right) \cosh (2 \pi  \alpha )- \left(c_g^2+1\right)} \log (N). \\
\end{split}
\ee 
For $D=2$, $\Delta S^{(n), c}_A$ do not depend on $\alpha$ and $N$ and they are given by (\ref{uiN})
\subsubsection*{Large $\alpha$ limit}
Here we would also like to study $\Delta S^{(n), c}_A$ in the large modular chemical potential limit ($\left|\alpha\right| \rightarrow \infty$) in stead of taking the large $N$ limit.

If we take the large $\alpha$ limit, $\Delta S^{(n), c}_A$ for $\left(\psi_a\right)^i_j$ vanish because the locally excited state reduces to a product state in the large $\alpha$ limit.
In $2$ dim. spacetime, if taking $|\alpha| \rightarrow \infty$ limit, $\Delta S^{(n), c}_A$ for $\Tr \bar{\psi}\psi$ reduce to 
\be
\Delta S^{(n), c}_A \sim 2\log{N},
\ee
where $n$ is an arbitrary integer. They are can be identified as (R$\acute{e}$nyi) entanglement entropy for a maximally entangled state in a certain charged sector.
In even higher dimension ($D>2$), if taking the same limit, $\Delta S^{(n), c}_A$ for $\Tr \bar{\psi}\psi$ reduce to
\be
\begin{split} 
&\Delta S^{(n\ge2), c}_A \sim\f{D}{2}\log{2}+2\log{N }+ \f{1}{1-n}\log{\left[\f{(1-c_g)^{2n}+(1+c_g)^{2n}}{2(1+c_g^2)^n}\right]} \\
&\Delta S^{c}_A\sim\f{D}{2}\log{2}+2\log{N }+\f{(1-c_g)^2}{2(1+c_g^2)}\log{\left(\f{(1+c_g^2)} {(1-c_g)^2}\right)}+\f{(1+c_g)^2}{2(1+c_g^2)}\log{\left(\f{(1+c_g^2)} {(1+c_g)^2}\right)}.
\end{split}
\ee

In $D=2$ case, $\Delta S^{(n), c}_A$ for $\Tr \psi^{\dagger}\psi$ do not depend on $\alpha$ and $N$ and it is given by (\ref{uiN}).
It is (R$\acute{e}$nyi) entanglement entropy for EPR state.
In even higher dimension, $\Delta S^{(n), c}_A$ for an arbitrary $n$ are given by
\be
\Delta S^{(n), c}_A\sim\f{D}{2}\log{2}+2\log{N}.
\ee
As explained earlier, they can be interpreted as a maximally entangled state in a certain charged sector.
Unlike the large $N$ limit, if taking the large $\alpha$ limit, entanglement in "quark" sector such as $\ket{\left(\bar{\psi}_a\right)^i_j}$ is partially enhanced. 
Even after taking the large $\alpha$ limit, von Neumann limit can be taken. Therefore the large $N$ limit is totally different from taking the large modular chemical potential limit.
In the large $\alpha$ limit in $D=2$, $\Delta S^{(n), c}_A$ for $\Tr\bar{\psi}\psi$ is greater than those for $\Tr\psi^{\dagger}\psi$. On the other hand, $\Delta S^{(n), c}_A$ for $\Tr\psi^{\dagger}\psi$ is greater than that for $\Tr\bar{\psi}\psi$ in the large $\left|\alpha\right|$ limit for $D>2$.
In the next subsection, let's interpret their dependence on $\alpha$ and $N$.

\subsubsection{Dependence on $N$ and $\alpha$}
We discuss the dependence of $\Delta S^{(n), c}_A$ on $N$ and $\alpha$. First we discuss it for $D=2$. The authors in \cite{m3} studied $\Delta S^{(n)}_A$ in the large $N$ free massless scalar field theory. For $n\ge 2$, $\Delta S^{(n)}_A$ is given by $\mathcal{O}(1)$ in the large $N$ limit. It can be interpreted as entanglement between the singlets (for example, $\ket{Tr \phi \phi}$) since contribution from entanglement in "quark" sector such as $\ket{\left(\phi\right)^{a}_b}$ is suppressed in the large $N$ limit. They found that the next leading correction is given by $\mathcal{O}(N^{2(n-1)})$. Therefore  we have to take the von Neumann limit $n \rightarrow 1$ before taking the large $N$ limit. 

However for $\Tr\bar{\psi} \psi$ and $\Tr \psi^{\dagger} \psi$ the von Neumann limit $n \rightarrow 1$ can be taken even after taking the large $N$ limit in the case of $D=2$. 
The components of the reduced density matrix for $\bar{\psi}\psi$ are given by $\ket{\left(\psi\right)^i_j}\bra{\left(\psi\right)^i_j}$ and $\ket{\left(\bar{\psi}\right)^i_j}\bra{\left(\bar{\psi}\right)^i_j}$. They depend on entanglement charge. Therefore its charge reduced density matrix is given by
\be
\begin{split}
\rho^{(n), c}_A =\f{1}{2N^2\cosh{2\pi \alpha}}\begin{pmatrix}
e^{2\pi \alpha}{\bf 1}_{N^2\times N^2} & 0 \\
0 & e^{-2\pi \alpha}{\bf 1}_{N^2\times N^2} \\
\end{pmatrix},
\end{split}
\ee 
where ${\bf 1}_{N^2\times N^2}$ is the $N^2\times N^2$ identity matrix.

There  are no singlet sectors. The effect of $N^2$ "quarks" is not suppressed in the large $N$ limit. Therefore even after taking  the large $N$ limit, the von Neumann limit can be taken. $\Delta S^{(n), c}_A$ for arbitrary $n$ is given by $\mathcal{O}(\log{N})$. It comes from entanglement in "quark" sector. Before taking the large $N$ limit, $\Delta S^{(n), c}_A$ depends on the entanglement chemical potential. However in the large $N$ limit the dependence of $\Delta S^{(n), c}_A$ on it disappears. 
Unless taking large $N$ limit,  $\Delta S^{(n)}_A$ ($\Delta S^{(n), c}_A$ without the modular chemical potential) is given by that for the maximally entangled state on the effective Hilbert space whose dimension is $2N^2$. If the large $N$ limit is taken, $\Delta S^{(n)}_A$ is given by that for the maximally entangled state on the effective Hilbert space whose dimension is $N^2$. If large $\alpha$ is added, $\Delta S^{(n), c}_A$ is given by $\Delta S^{(n)}_A$ for the maximally entangled state on the effective Hilbert space whose dimension is $N^2$.

The components of the reduced density matrix for $\Tr\psi^{\dagger}\psi$ is given by $\ket{0}\bra{0}$ and $\ket{\Tr \psi^{\dagger}\psi}\bra{\Tr \psi^{\dagger}\psi}$. They do not depend on entanglement charge. Therefore its charged reduced density matrix is given by
\be
\begin{split}
 \rho^c_A =\f{1}{2}
\begin{pmatrix}
1 & 0 \\
0 & 1 \\
\end{pmatrix}.
\end{split}
\ee 

Therefore $\Delta S^{(n), c}_A$ for arbitrary $n$ is given by (R$\acute{e}$nyi) entanglement entropy for an EPR state. It dose not depend on $\alpha$ and $N$.  In the $2$ dimensional free massless fermionic field theory with $U(N)$ symmetry, $\Delta S^{c}_A$ for $\Tr\bar{\psi}\psi$ is greater than that for $\Delta S^{c}_A$ for $\Tr\psi^{\dagger}\psi$.

 Although the von Neumann limit can be taken in the $2$ dimensional free massless fermionic field theory, we can not taking the von Neumann limit ($n\rightarrow 1$) after taking large $c$ limit in the specific interacting theories even in $D=2$ as in \cite{m3}.
 
For $D>2$, the components of the reduced density matrices for $\Tr \bar{\psi}\psi$ and $\Tr \psi^{\dagger} \psi$ necessarily include those in the singlet sector and quark sectors. Entanglement in quark sector is suppressed in the large $N$ limit. Therefore the von Neumann limit $n \rightarrow 1$ has to be taken before taking the large $N$ limit when we compute $\Delta S^{c}_A$ in the large $N$ limit. In this case the effect of modular chemical potential on $\Delta S^{(n), c}_A$ does not disappear even in the large $N$ limit.

Finally we would like to comment on the difference between the effect of $N$ and $\alpha$ on $\Delta S^{(n), c}_A$. As well as we study, the modular chemical potential enhances the effect of entanglement in quark sector in the large $\alpha$ limit. On the other hand, the parameter $N$ enhances that of entanglement in the singlet sector for $\Delta S^{(n\ge2), c}_A$ and $\Delta S^{(n\ge2)}_A$ in $D\neq2$. After taking the large $N$ limit, the von Neumann limit can not be taken. On the other hand, even after taking the large $\alpha$ limit, von Neumann limit can be taken because there are not $\mathcal{O}(e^{2(n-1)\pi \alpha})$ correction. As in (\ref{nbp}) and (\ref{ndp}), in the large $N$ limit $\Delta S^{c}_A$ for $\Tr\bar{\psi}\psi$ is greater than that for $\Tr\psi^{\dagger}\psi$ in $D>2$. On the other hand, in the large modular chemical potential limit, $\Delta S^{(c)}_A$ for $\Tr \psi^{\dagger}\psi$ is greater than that for $\Tr\bar{\psi} \psi$\footnote{In $D=2$ $\Delta S^{(n), c}_A$ for $\Tr \bar{\psi} \psi$ is greater than that for $\Tr \psi^{\dagger}\psi$ in both limits.}. In this sense, taking the large $N$ limit is totally different from taking the large $|\alpha|$ limit. 
    
In the end of this section we study the time evolution of $\Delta S^{(n), c}_A$ in $4$ dimensional space time without $U(N)$ symmetry.
\subsection{Time evolution of CREEs}
In this subsection, we study the time evolution of  $\Delta S^{(n), c}_A$ in $4$ dimensional space time without $U(N)$ symmetry. Here for simplicity we consider $\Delta S^{(2), c}_A$ for $\psi_a$, $\bar{\psi} \psi$ and $\psi^{\dagger} \psi$. 
The time evolution of $\Delta S^{(2), c}_A$ for $\psi_a$, $\bar{\psi} \psi$ and $\psi^{\dagger} \psi$ is plotted in Figures 8 and 9 and 10 respectively. We especially consider the behavior of $\Delta S^{(2), c}_A$ for $\bar{\psi}\psi$ and $\psi^{\dagger}\psi$. When $\alpha$ is small ($\left|\alpha\right| \sim 0$) or large ($\left|\alpha\right|\rightarrow \infty$), it can be interpreted physically.

In the former case, the time evolution of $\Delta S^{(2), c}_A$ for $\bar{\psi}\psi$,  and $\psi^{\dagger}\psi$ is similar to that of $\Delta S^{(2)}_A$ for them.
Therefore it can be interpreted similarly to $\Delta S^{(2), c}_A$ without the modular chemical potential. Entanglement in all sectors can contribute to $\Delta S^{(2), c}_A$.

On the other hand, in the later case, entanglement between (anti-)particles in a certain charged sector can only contribute to $\Delta S^{(2), c}_A$. Here we assume that $\g^t\g^1$ is given by $diag(1,-1,1,-1)$ for simplicity and also assume that even unless taking late time limit, the reduced density matrices for $\Delta S^{(2), c}_A$ can be written by 
\be
\begin{split}
\rho^{bp}_A&=P_1\ket{0}_A\bra{0}_A+P_2\ket{\bar{\psi}\psi}_A\bra{\bar{\psi}\psi}_A+P_3\left(\ket{\bar{\psi}_1}\bra{\bar{\psi}_1}+\ket{\bar{\psi}_3}\bra{\bar{\psi}_3}\right)+P_5\left(\ket{\bar{\psi}_2}\bra{\bar{\psi}_2}+\ket{\bar{\psi}_4}\bra{\bar{\psi}_4}\right) \\
&+P_4\left(\ket{\psi_1}\bra{\psi_1}+\ket{\psi_3}\bra{\psi_3}\right)+P_6\left(\ket{\psi_2}\bra{\psi_2}+\ket{\psi_4}\bra{\psi_4}\right) \\
\rho^{dp}_A&=\bar{P}_1\ket{0}_A\bra{0}_A+\bar{P}_2\ket{\phi\psi}_A\bra{\phi\psi}_A+\bar{P}_3\left(\ket{\phi_1}\bra{\phi_1}+\ket{\phi_3}\bra{\phi_3}\right)+\bar{P}_5\left(\ket{\phi_2}\bra{\phi_2}+\ket{\phi_4}\bra{\phi_4}\right) \\
&+\bar{P}_4\left(\ket{\psi_1}\bra{\psi_1}+\ket{\psi_3}\bra{\psi_3}\right)+\bar{P}_6\left(\ket{\psi_2}\bra{\psi_2}+\ket{\psi_4}\bra{\psi_4}\right) \\
\end{split}
\ee
where $\rho^{bp, c}_A$ and $\rho^{dp, c}_A$ respectively correspond to the charged reduced density matrices for  $\bar{\psi}\psi$ and $\psi^{\dagger}\psi$. We also assume that $\sum_{i}P_i=\sum_i\bar{P}_i =1$. In both case, the behavior of $\Delta S^{(2), c}_A$ can be interpreted in the same way. Therefore we consider $\Delta S^{(2), c}_A$ for only $\bar{\psi}\psi$.
 If the limit $\alpha \rightarrow \infty$ is taken, entanglement between two components $\bar{\psi}_a$ can only contribute to $\Delta S^{(2), c}_A$ for $\bar{\psi}\psi$ at $t=l$. Therefore although only $P_1$ does not vanish before $t=l$, only $P_3$ or $P_5$ can be none-zero at $t=l$. Because we assume that $\g^t\g^1=diag(1,-1,1,-1)$, only one of them does not vanish at $t=l$. Then $\Delta S^{(2), c}_A$ suddenly increases at $t=l$ as in $2$d CFT.  $\Delta S^{(2), c}_A$ is given by $\log{2}$.  However in this case entanglement between the residual two components $\bar{\psi}_a$ can contribute to $\Delta S^{(2), c}_A$ after $t=l$. Therefore $\Delta S^{(2), c}_A$ keeps to increase after $t=l$ and approaches the late values in the limit $\alpha \rightarrow \infty$, which are obtained in the previous subsection. The time evolution of $\Delta S^{(2), c}_A$ for $\psi^{\dagger}\psi$ can be interpreted in the same way.

The time evolution of $\Delta S^{(2)}_A$ outside the region where we study is interesting because the structure of entanglement between all sectors is deformed by charge chemical potential. The effect of entanglement in some sector is partially enhanced and that of entanglement in other sector is partially suppressed. Then the effect of the modular chemical potential can compete with the original features of operators.  The Figure 9 and 10 show that $\Delta S^{(n), c}_A$ can have a peak in this region. 
To study it is one of the important future problems in order to study features of charged (R$\acute{e}$nyi) entanglement entropy.

\begin{figure}[h!]
  \centering
  \includegraphics[width=8cm]{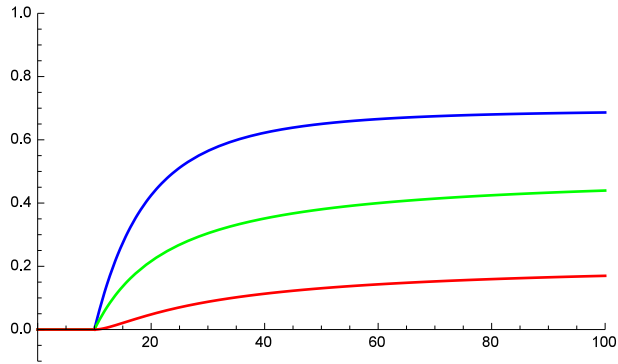}
  \caption{The plot of $\Delta S^{(2), c}_A$ for real entanglement chemical potential $\alpha$. The horizontal axis corresponds to time $t$. The vertical line corresponds to $\Delta S^{(2), c}_A$.  $\psi_a$ is the local operator which acts on the ground state. Here $l=10, \alpha =\f{1}{2\pi}$. The blue, green and red lines correspond to $\left(\g^t\g^1\right)_{aa} = -1, 0$ and $1$.}
\end{figure}
 \begin{figure}[h!]
  \centering
  \includegraphics[width=8cm]{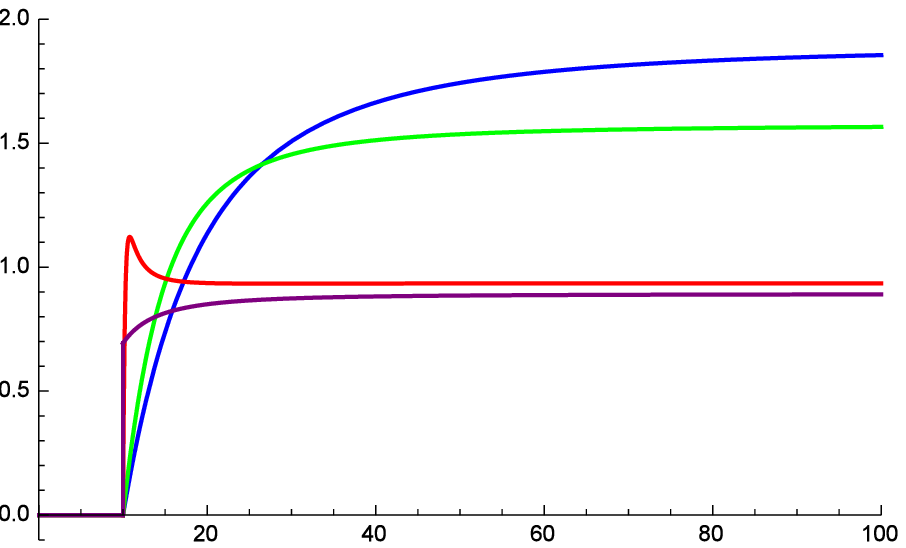}
  \caption{The plot of $\Delta S^{(2), c}_A$ with real entanglement chemical potential $\alpha$. The horizontal axis corresponds to time $t$. The vertical line corresponds to $\Delta S^{(2), c}_A$.  $\bar{\psi}\psi$ is the local operator which acts on the ground state. Here $l=10$. The blue, green and red lines correspond to $\alpha=0, \f{1}{2\pi}$ and $\f{2}{\pi}$. The purple line corresponds to $\alpha \rightarrow \infty$.}
\end{figure}
\begin{figure}[h!]
  \centering
  \includegraphics[width=8cm]{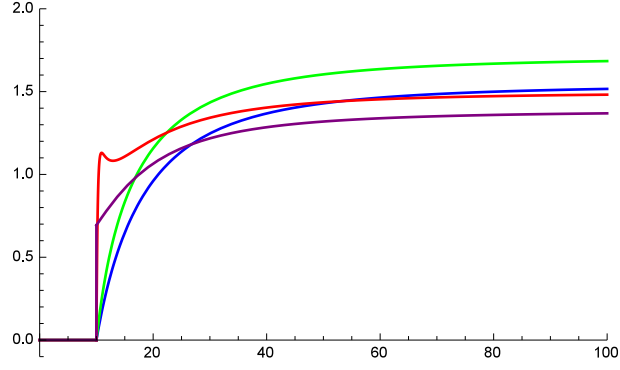}
  \caption{The plot of $\Delta S^{(2), c}_A$ with real entanglement chemical potential $\alpha$. The horizontal axis corresponds to time $t$. The vertical line corresponds to $\Delta S^{(2), c}_A$.  $\psi^{\dagger}\psi$ is the local operator which acts on the ground state. Here $l=10$. The blue, green and red lines correspond to $\alpha=0, \f{1}{2\pi}$ and $\f{2}{\pi}$.The purple line corresponds to $\alpha \rightarrow \infty$}
\end{figure}

\section{Holographic results for $d\ge 4$ }

In this section, following \cite{m3}, we compute the excess in charged (R$\acute{e}$nyi) entanglement entropies at large central charge and in the limit of a small chemical potential. We consider a charged topological black hole in $AdS_d$ with $d\ge 4$ \cite{tpbh} (see also \cite{chm}). At large central charge, we can compute the $2n$-point correlation functions that appear in the the R\'enyi entropies from holography \cite{m3}. More precisely, at large $c$, the $2n$-point function factorizes into a product of the two-point functions that can be approximated by the exponent of the geodesic length computed in the charged topological black hole background (with $\Sigma_n$ at the boundary). Below, we follow \cite{m3} very closely and only stress the differences that appear for the charged case.\\
The metric of the (Euclidean) charged topological black hole in $AdS_{d+1}$ is given by
\be
ds^2=f(r)d\tau^2+\frac{dr^2}{f(r)}+r^2d\phi^2+r^2e^{-2\phi}dx^2_i,\label{MET}
\ee
where 
\be
f(r)=-1-\frac{m}{r^{d-2}}+\frac{r^2}{R^2}+\frac{q^2}{r^{2(d-2)}}.
\ee
In addition, the solution contains a bulk gauge field
\be
A=\left(\sqrt{\frac{2(d-1)}{(d-2)}}\frac{Rq}{R_*l_*r^{d-2}}-\frac{\mu}{2\pi R_*}\right)id\tau,
\ee
that vanishes at the horizon $r_+$. This gives a relation between $\mu$ and the charge $q$
\be
\frac{q^2}{r^{2(d-2)}_+}=\left(\sqrt{\frac{d-2}{2(d-1)}}\frac{\mu l_*}{2\pi R}\right)^2\equiv\mu^2_d,
\ee
where for simplicity of notation we defined $\mu^2_d$. Once we express the mass $m$ and $f(r)$ in terms of $\mu^2_d$, the temperature can be written as
\bea
\beta^{-1}=T_{H}=\left.\frac{1}{4\pi}\partial_r f(r)\right|_{r_+}=\frac{r_+d}{4\pi R^2}-\frac{d-2}{4\pi r_+}\left[1+\mu^2_d\right],
\eea
where $\beta =2n \pi R$.
Then, we have the radius 
\be
r_{+}=\frac{R}{dn}+ R\sqrt{1-\frac{2}{d}+\frac{1}{d^2n^2}+\frac{d-2}{d}\mu^2_d}.\label{rpnd}
\ee
Putting all together, we can express the mass in terms of $n$, $d$ and $\mu$ 
\be
m=2R^{d-2}\left(\frac{1}{dn}+\sqrt{()}\right)^{d-2}\left(\frac{1}{d^2n^2}-\frac{1}{d}+\frac{d-1}{d}\mu^2_d+\frac{1}{dn}\sqrt{()}\right)\equiv R^{d-2}F_{(d,n,\mu)},
\ee
where $\sqrt{()}$ stands for the square root in \eqref{rpnd}. Notice that  due to the presence of the chemical potential, now $m(n=1)\neq 0$ (for $\mu^2_d=0$, $m(1,0)=0$ as before \cite{m3}).\\
Analogously to \cite{m3}, we can compute the length of the geodesics in metric \eqref{MET} at fixed $x_i$
\be
\frac{L}{2}=\int^{r_\Lambda}_{r_*}\sqrt{f(r)\tau'^2+\frac{1}{f(r)}+r^2\phi'^2}\,dr\equiv\int^{r_\Lambda}_{r_*}\LL\,dr.
\ee
The corresponding equations of motion are
\bea
\frac{f(r)\tau'}{\LL}=C_1,\qquad
\frac{r^2\phi'}{\LL}=C_2,
\eea
where $C_{1}$ and $C_2$ are arbitrary constants. Using the divergence of the derivatives at the turning point $r_*$ we can further eliminate one constant, so that the distances between points are 
\bea
\Delta\tau\equiv\tau_f-\tau_i&=&2\int^{\infty}_{r_*}\frac{r\sqrt{f(r_*)(r^2_*-C^2_2)}\,dr}{f(r)\sqrt{f(r)r^2_*(r^2-C^2_2)-f(r_*)r^2(r^2_*-C^2_2)}}\\
\Delta\phi\equiv\phi_f-\phi_i&=&2\int^{\infty}_{r_*}\frac{C_2r_*\,dr}{r\sqrt{f(r)r^2_*(r^2-C^2_2)-f(r_*)r^2(r^2_*-C^2_2)}},\label{INT}
\eea
and the geodesic length is
\be
\frac{L}{2}=r_*\int^{r_\Lambda}_{r_*}\frac{r\,dr}{\sqrt{f(r)r^2_*(r^2-C^2_2)-f(r_*)r^2(r^2_*-C^2_2)}}.\label{Length}
\ee
We are interested in the late time behavior of the correlation function for which the distances between points become \cite{m3}
\be
\frac{\Delta\tau}{R}=\pi+O(t^{-2}),\qquad \Delta\phi=i\left(\pi-\frac{2\varepsilon}{t}\right)+O(t^{-2}),\label{Ltime}
\ee
we can then set $C^2_2=-|C_2|^2$ and define
\be
\beta=\frac{R|C_2|}{r_*}.
\ee
The relevant answer in the late time limit can be extracted by expanding the integrands for large $r_*$ as well as keeping the leading term in large $\beta$. The length of the geodesic in this limit is given by
\be
\frac{L}{2R}\simeq \log\left(\frac{2r_\Lambda}{r_*\beta}\right)+O(r^{-1}_\Lambda).\label{Lengthlt}
\ee
Finally, we just have to determine the relation between $\beta$ and $m(n,\mu)$. The algorithm is that we focus on the difference between $|\Delta\phi|$ and $\frac{\Delta \tau}{R}$ by performing the expansion of the two integrals in the inverse of $r_*$ and integrating to the leading term in $1/\beta$ and also to the first order in the chemical potential.\\ 
For example in 4 dimensions $(d=3)$ this leads to the relation
\bea
\frac{2\varepsilon}{t}\simeq \frac{R^2}{r^2_*\beta}-\frac{mR^2}{3r^3_*}+O(r^{-4}_{*}),
\eea
and in order for the two expressions on the right hand side to be of the same order, we must have
\be
\beta\sim\alpha_3\frac{r_*}{m},
\ee
where $\alpha_3$ is an $O(1)$ number. From that we can fix 
\be
\frac{2\varepsilon}{t}\sim\frac{R^2m}{r^3_*}c_3,
\ee
with some $O(1)$ constant $c_3$, so we can express $r_*$ in \eqref{Lengthlt} in terms of the relevant parameters.\\
Analogous analysis in $d$-dimensions yields
\be
\frac{2\epsilon}{t}\sim \frac{R^2}{r^2_*\beta}-\lambda_{d}\frac{R^2m}{r^d_*}=c_d\frac{R^2m(n,\mu)}{r^d_*},
\ee
and the length of the geodesic becomes
 \be
\frac{L^{(n)}}{2R}\sim\log\left(\frac{2r_\Lambda}{r_*\beta}\right)
=\log\left(\frac{2r_\Lambda m(n,\mu)}{\alpha_d\, r^{d-1}_*}\right).
\ee

The above geodesic gives us the propagator on $\Sigma_n$ in $d$-dimensions and at large $c$ from holography
\be
e^{-\frac{2\Delta_O}{R} L^{(n)}}\sim\left(\frac{\alpha_dR^{\frac{d-2}{d}}}
{\frac{2r_\Lambda}{R}m(n,\mu)^{\frac{1}{d}}}
\left(\frac{2\epsilon}{c_dt}\right)^{\frac{1}{d}-1}\right)^{4\Delta_O}.
\ee
This is our main holographic result from which we can now evaluate the correlators in the large $c$ limit.\\
There are two allowed sets of Wick contractions that we can take (see detailed discussion in \cite{m1}) and the change in the n-th CREE becomes
\be
\Delta S^{(n)}_A\simeq\frac{4n\Delta}{d(n-1)}\log\left(\frac{F_{(d,n,\mu)}\,t}{2\,\epsilon}\right)+C_{(n,d)}-\frac{1}{n-1}\log 2,
\ee
where $C_{(n,d)}$ is some non-universal constant. This is the same form as for the uncharged case but now $F_{(d,n,\mu)}$ depends on the chemical potential. Clearly, holographically the effect of a small charge is only seen in the details of the logarithmic growth with time but the coefficient of the logarithm is unchanged. Note however that this answer is based on the expansion in large $r_*$, so that we can perform the very complicated integrals in arbitrary dimension. As we will see in the Appendix D, for $d=2$ one gets more analytic control and the expansion is not needed.

\subsubsection*{Discussion}
Here we assumed that the real modular chemical potential is added. In our analysis, $\Delta S^{(n \ge 2)}_A$ does not approach to a constant even at the late time $(t \gg l \gg \epsilon)$. This behavior might be interpreted as follows. We implicitly assume that the modular chemical potential $\alpha$ is very small and $\beta$ and $r_*$ are large. Under this assumption, we evaluate $\Delta \tau$, $\Delta \phi$ and $L$ by expanding with respect to $r_*$ and $\beta$.  On the other hand, the black hole horizon ($r_+$) depends on the modular chemical potential as in (\ref{rpnd}) because $\mu \sim \alpha$. Therefore, if $\alpha$ is large, the black hole horizon can be large and geodesics, that compute the correlator holographically, can probably attach to it. Hence, our analysis is expected to break down in the limit of a large modular chemical potential\footnote{The results in this section is expected to be reliable when $\f{\epsilon}{t}$ is much greater than the modular chemical potential.}. Therefore, it is possible that $\Delta S^{(n), c}_A$ approaches a constant at the very late time limit for sufficiently large chemical potentials. Actually, as in Appendix D, $\Delta S^{(n), c}_A$ is given by constant in the $AdS_3$ example\footnote{However, since we are not certain about the physical interpretation of that behavior, we have added the example to Appendix D. Certainly, understanding of this setup and behavior is one of the important future problems that can help us to explore the new features of charged (R$\acute{e}$nyi) entanglement entropies.}.

\section{Conclusions and future problems} 
Here, we finish by brief conclusions and outline some future problems. In this work we focused on the excesses of (R$\acute{e}$nyi) and charged (R$\acute{e}$nyi) entanglement entropies after excitations by local fermionic operators. 
We derived the dependence of the coefficient $c_g$ of $\g^t \g^1$ on spacetime dimension $D$ (given by \eqref{dcog}) and fixed the algebra of the fermionic quasi-particles. The coefficient can be interpreted as follows. For $D=2p~ (p \in \mathbb{Z})$, fermionic field can be written as a linear combination of fields labeled by eigenvalues of p spins and one of them can be chosen as that of $\g^t \g^1$.  Fermions' propagation depends on the representation of $\g^t \g^1$ and $c_g$ depends on the number of $p-1$ spins except of $\g^t\g^1$. Fermions which have the same number of negative eigenvalues contribute equally to probability of the fermionic (anti-)particles to be in $A$ or $B$.  Therefore, the square of a combination matrix appears in $c_g$. If we choose a more complicated subsystem such as $x^1 \ge 0$ and $x^2 \ge 0$ as the subsystem A, $c_g$ is expected to depend on other generators.  

We have also considered the excesses of charged (R$\acute{e}$nyi) entanglement entropy for locally excited states in the free massless fermionic field theory. We found that the late values of $\Delta S^{(n), c}_A$ can be affected by the modular chemical potential. For $\bar{\psi}\psi$, $\Delta S^{c}_A$ in $4d$ monotonically decreases when $\left|\alpha\right|$ increases and approaches entanglement entropy for a certain charged sector. For $\psi^{\dagger}\psi$, $\Delta S^{c}_A$ in $4d$ does not monotonically decreases when $\left|\alpha\right|$ increases. Until the certain value $\left|\alpha_m\right|$ of $\left|\alpha\right|$, it increases. After $\left|\alpha\right|$ is over $\left|\alpha_m\right|$, they decrease and approach (R$\acute{e}$nyi) entanglement entropies for the maximally entangled state in a certain charged sector. If we recognize $\Delta S^{(n),c}_A$ as the amount of quantum information, the results indicate that it can be controlled by changing the modular chemical potential. The way of increasing or decreasing depends on the local operator. If we add the large amount of $\left|\alpha\right|$ to the reduced density matrix for $\bar{\psi} \psi$ and $\psi^{\dagger}\psi$, we can obtain  (R$\acute{e}$nyi) entanglement entropies in a certain charged sector. In $D=4$,  $\Delta S^{c}_A$ without the modular chemical potential for $\bar{\psi}\psi$ is greater than that for $\psi^{\dagger}\psi$. On the other hand, in the large modular chemical potential limit, that for  $\bar{\psi}\psi$ is smaller than that for $\psi^{\dagger}\psi$. We have also studied the time evolution of $\Delta S^{(n)}_A$. In the small and large $\alpha$ region we found that it can be interpreted physically. Especially, $\Delta S^{(2), c}_A$ for $\bar{\psi}\psi$ and $\psi^{\dagger}\psi$ suddenly increases at $t=l$ in the large modular chemical potential limit.
They increase and  approach a certain constants at the late time. We found that this behavior can be interpreted in terms of quasi-particles.

We also studied $\Delta S^{(n)}_A$ and $\Delta S^{(n), c}_A$ in the $D$ dimensional free massless fermionic field theory with $U(N)$ symmetry with $D$ given by even integer. We analyzed the difference between taking the large $N$ limit and taking the large modular chemical potential limit. In the large $N$ limit, the entanglement in singlet sector is enhanced. After taking the large $N$ limit, the von Neumann limit can not be taken for $D>2$. In $D=2$, the limit $n \rightarrow 1$ can be taken for $\Delta S^{(n), c}_A$ for $\Tr\bar{\psi}\psi$ and $\Tr \psi^{\dagger}\psi$ after taking the large $N$ limit since there is entanglement in the singlet or "quark" sector. On the other hand, in the large $\alpha$ limit, the entanglement in a certain charge sector is enhanced.   After taking the large $\alpha$ limit, the von Neumann limit can be taken for any even $D$.

Finally, we studied $\Delta S^{(n), c}_A$ in the holographic field theory. In this case, even at the late time limit, $\Delta S^{(n), c}_A$ grow logarithmically with time. In our analysis, the coefficient of logarithmic term dose not depend on the modular chemical potential. However, the detailed time evolution of $\Delta S^{(n \ge 2), c}_A$ depends on the modular chemical potential. We implicitly assumed that the modular chemical potential is very small.  When the modular chemical potential is added to the reduced density matrix, the black hole horizon of its gravity dual depends on $\alpha$. If the large modular chemical potential is added, the black hole horizon is large. Then geodesics can probably attach its horizon.
Therefore,  $\Delta S^{(n), c}_A$ can probably be constant as in Appendix D.

\medskip
There are many interesting future problems and we list a few below: 
\begin{itemize}
\item{It is interesting to study the dependence of $\Delta S^{(n),c}_A$ on spacetime when we take different shapes of the subsystem A.}

\item{We have studied time evolution of $\Delta S^{(n),c}_A$ for small and large real modular chemical potential. It would be interesting to understand their behaviour in the intermediate regions too.}

\item{Understanding $\Delta S^{(n), c}_A$ with more complicated charges or many charges simultaneously (integrable models) is one of the future problems.}

\item{It would also be very interesting to categorize local operators by the properties of $\Delta S^{(n), c}_A$}.

\item{It is also an important problem to study $\Delta S^{(n), c}_A$ in the holographic field theories in higher dimensions for large chemical potential and see if that leads to the constant behaviour at late times.}

\end{itemize}

\subsection*{Acknowledgements}
We would like to thank Shunji Matsuura and  Mitsutoshi Fujita for collaboration in the early stage of this work and Alvaro Veliz Osorio for comments on the draft and discussions.
We would like to thank Tadashi Takayanagi for helpful discussion. TN is supported by JSPS fellowship. PC is supported by  the Swedish Research Council (VR) grant 2013-4329.

\appendix
\def\thesection{ \Alph{section}}

\section{Propagators}

\subsection*{Propagators for twisted fields}
After performing the analytic continuation in (\ref{ac}), we take the limit $\epsilon \rightarrow 0$. In this limit, some of the twisted Green function dominantly contribute to $\Delta S^{(n), c}_A$. We call them dominant Green functions which are listed in Table 11 and 12.
\begin{table}[htb]
  \begin{center} 
    \begin{tabular}{|c|c|} \hline
      $n=1,~~t\le l$ & $$     \\ \hline
      $S_{ab}(\theta-\theta_2)=-S_{ab}( \theta_2-\theta) $ &  $i\gamma^t_{ab}\left(-\frac{1}{16 \pi ^2 \epsilon ^3}\right) $  \\ \hline \hline
    $t > l$   &     \\ \hline
     $S_{ab}(\theta-\theta_2)$ & $ i\left[\left(-\frac{e^{-i 2 \pi  \alpha_E } (t-l)+l+t}{32 \pi ^2 t \epsilon ^3}\right)\gamma^t_{ab}+\frac{\left(-e^{-i 2 \pi  \alpha_E } +1\right) (l-t) (l+t)}{64 \pi ^2 t^2 \epsilon ^3}\g^1_{ab}\right]$  \\ \hline
$S_{ab}(\theta_2-\theta) $ & $ i\left[\left(\frac{e^{i 2 \pi  \alpha_E } (t-l)+l+t}{32 \pi ^2 t \epsilon ^3}\right)\gamma^t_{ab}+\frac{\left(e^{i 2 \pi  \alpha_E } -1\right) (l-t) (l+t)}{64 \pi ^2 t^2 \epsilon ^3}\g^1_{ab}\right]$  \\ \hline \hline
$n\ge2,~~t\le l$ &      \\ \hline
      $S_{ab}(\theta-\theta_2)=-S_{ab}(\theta_2-\theta) $ &  $i\gamma^t_{ab}\left(-\frac{1}{16 \pi ^2 \epsilon ^3}\right) $  \\ \hline \hline
  $t > l$      &    \\ \hline
     $S_{ab}(\theta-\theta_2)=-S_{ab}(\theta_2-\theta)$ & $ -i\frac{l+t}{32 \left(\pi ^2 t\right) \epsilon ^3}\gamma^t_{ab}+i\left(\frac{(l-t) (l+t)}{64 \pi ^2 t^2 \epsilon ^3}\right)\gamma^1_{ab}$  \\ \hline
$S_{ab}( \theta-\theta_2-2(n-1)\pi) $ & $i(-1)^{n+1}e^{-i2n\pi \alpha_E}\left[\left(\frac{ (l-t)}{32 \pi ^2 t \epsilon ^3}\right)\gamma^t_{ab}-\left(\frac{ (l-t) (l+t)}{64 \pi ^2 t^2 \epsilon ^3}\right)\gamma^1_{ab}\right]$  \\ \hline
$S_{ab}(\theta_2-\theta+2(n-1)\pi) $ & $i(-1)^{n}e^{i2n\pi \alpha_E}\left[\left(\frac{ (l-t)}{32 \pi ^2 t \epsilon ^3}\right)\gamma^t_{ab}-\left(\frac{ (l-t) (l+t)}{64 \pi ^2 t^2 \epsilon ^3}\right)\gamma^1_{ab}\right]$  \\ \hline
$S_{ab}(\theta-\theta_2+2\pi)=-S_{ab}( \theta_2-\theta-2\pi)$ & $ i\frac{l-t}{32 \pi ^2 t \epsilon ^3}\gamma^t_{ab}+i\gamma^1_{ab}\left(-\frac{(l-t) (l+t)}{64 \left(\pi ^2 t^2\right) \epsilon ^3}\right)$  \\ \hline
    \end{tabular}
  \end{center}
  \caption{The table of green function $S_{ab}(\theta_{\alpha}-\theta_{\beta})$ for twisted fields in $4$ dim..}
\end{table}
If we define 
\be
\begin{split}
-\left\langle\mathcal{T} \psi_a(r, \theta)\psi^{\dagger}_b(r_2, \theta_2)\right\rangle=V_{ab}(\theta-\theta_2)=i S_{ac}(\theta-\theta_2)\gamma^t_{cb}.
\end{split}
\ee
In the limit $\epsilon \rightarrow 0$, some of $V_{ab}$ can contribute dominantly and they are listed in Table 11 and 12.
\begin{table}[h!]
  \begin{center} 
    \begin{tabular}{|c|c|} \hline
      $n=1,~~t\le l$ & $$     \\ \hline
      $V_{ab}(\theta_2-\theta) =-V_{ab}(\theta-\theta_2)$ &  ${\bf 1}_{ab}\left(-\frac{1}{16 \pi ^2 \epsilon ^3}\right) $  \\ \hline \hline
    $t > l$   &     \\ \hline
     $V_{ab}(\theta-\theta_2)$ & $ -\left[\left(\frac{e^{-i 2 \pi  \alpha_E } (t-l)+l+t}{32 \pi ^2 t \epsilon ^3}\right){\bf 1}_{ab}-\frac{\left(1-e^{-i 2 \pi  \alpha_E }\right) (l-t) (l+t)}{64 \pi ^2 t^2 \epsilon ^3}\left(\g^t\g^1\right)_{ab}\right]$  \\ \hline
$V_{ab}(\theta_2-\theta)$ & $-\left[\left(-\frac{e^{i 2 \pi  \alpha_E } (t-l)+l+t}{32 \pi ^2 t \epsilon ^3}\right){\bf 1}_{ab}-\frac{\left(e^{i 2 \pi  \alpha_E } -1\right) (l-t) (l+t)}{64 \pi ^2 t^2 \epsilon ^3}\left(\g^t\g^1\right)_{ab}\right]$  \\ \hline \hline
$n\ge2,~~t\le l$ &      \\ \hline
      $V_{ab}(\theta_2-\theta) =-V_{ab}(\theta-\theta_2)$ &  ${\bf 1}_{ab}\left(-\frac{1}{16 \pi ^2 \epsilon ^3}\right) $  \\ \hline \hline
  $t > l$      &    \\ \hline
     $V_{ab}(\theta-\theta_2)=-V_{ab}(\theta_2-\theta)$ & $ -\frac{l+t}{32 \left(\pi ^2 t\right) \epsilon ^3} {\bf 1}_{ab}-\left(\frac{(l-t) (l+t)}{64 \pi ^2 t^2 \epsilon ^3}\right)\left(\gamma^1\gamma^t\right)_{ab}$  \\ \hline
$V_{ab}(\theta-\theta_2-2(n-1)\pi)$ & $(-1)^{n}e^{-i2n\pi \alpha_E}\left[\left(-\frac{ (l-t)}{32 \pi ^2 t \epsilon ^3}\right){\bf 1}_{ab}-\left(\frac{ (l-t) (l+t)}{64 \pi ^2 t^2 \epsilon ^3}\right)\left(\gamma^1\gamma^t\right)_{ab}\right]$  \\ \hline
$V_{ab}(\theta_2-\theta+2(n-1)\pi) $ & $(-1)^{n+1}e^{i2n\pi \alpha_E}\left[\left(-\frac{ (l-t)}{32 \pi ^2 t \epsilon ^3}\right){\bf 1}_{ab}-\left(\frac{ (l-t) (l+t)}{64 \pi ^2 t^2 \epsilon ^3}\right)\left(\gamma^1\gamma^t\right)_{ab}\right]$  \\ \hline
$V_{ab}(\theta-\theta_2+2\pi)=-V_{ab}( \theta_2-\theta-2\pi)$ & $\frac{l-t}{32 \pi ^2 t \epsilon ^3}{\bf 1}_{ab}-\left(\gamma^1\gamma^t\right)_{ab}\left(-\frac{(l-t) (l+t)}{64 \left(\pi ^2 t^2\right) \epsilon ^3}\right)$  \\ \hline
    \end{tabular}
  \end{center}
   \caption{The table of green function $V_{ab}(\theta_{\alpha}-\theta_{\beta})$ for twisted fields in $4$ dim. .}
\end{table}
\subsection*{Propagators for untwisted fields}
Here we list up untwisted dominant propagators in $D=6, 8$ in Table 13 and 14 after the analytic continuation in (\ref{ac}).
\begin{table}[h!]
  \begin{center} 
   \scalebox{0.9}{  
    \begin{tabular}{|c|c|} \hline
      $D=6,~~n=1$ & $$     \\ \hline
      $S_{ab}( \theta-\theta_2)=-S_{ab}( \theta_2-\theta) $ &  $-i\frac{1}{32 \pi ^3 \epsilon ^5}\gamma^{t}_{ab} $  \\ \hline
$V_{ab}( \theta-\theta_2)=-V_{ab}(\theta_2-\theta) $ &  $-\frac{1}{32 \pi ^3 \epsilon ^5}{\bf 1}_{ab}$  \\ \hline \hline
$n\ge2,~~t\le l$ &      \\ \hline
      $S_{ab}(\theta-\theta_2)=-S_{ab}(\theta_2-\theta) $ &  $-i\frac{1}{32 \pi ^3 \epsilon ^5}\gamma^{t}_{ab}$  \\ \hline
$V_{ab}(\theta-\theta_2)=-V_{ab}(r_2, r, \theta_2-\theta) $ &  $-\frac{1}{32 \pi ^3 \epsilon ^5}{\bf 1}_{ab}$  \\ \hline \hline
  $t > l$      &    \\ \hline
     $S_{ab}(\theta-\theta_2)=-S_{ab}(\theta_2-\theta)$ & $ i\frac{(l-2 t) (l+t)^2}{128 \pi ^3 t^3 \epsilon ^5}\gamma^t_{ab} -i\frac{3 \left(l^2-t^2\right)^2}{512 \pi ^3 t^4 \epsilon ^5}\gamma^1_{ab}$  \\ \hline
$S_{ab}(\theta-\theta_2-2(n-1)\pi)=-S_{ab}(\theta_2-\theta+2(n-1)\pi) $ & $(-1)^{n}\left[i\frac{(l-t)^2 (l+2 t)}{128 \pi ^3 t^3 \epsilon ^5}\gamma^t_{ab}-i\frac{3 \left(l^2-t^2\right)^2}{512 \pi ^3 t^4 \epsilon ^5}\gamma^1_{ab}\right]$  \\ \hline
$S_{ab}(\theta-\theta_2+2\pi)=-S_{ab}( \theta_2-\theta-2\pi)$ & $-i\frac{(l-t)^2 (l+2 t)}{128 \pi ^3 t^3 \epsilon ^5}\gamma^t_{ab}+i\frac{3 \left(l^2-t^2\right)^2}{512 \pi ^3 t^4 \epsilon ^5}\gamma^1_{ab}$  \\ \hline
     $V_{ab}(\theta-\theta_2)=-V_{ab}(\theta_2-\theta)$ & $ \f{(t+l)^2}{128\pi^3t^3\ep^5}\left[(l-2t){\bf 1}_{ab}+\f{3(t-l)^2}{4t}\left(\gamma^1\gamma^t\right)_{ab}\right]$  \\ \hline
$V_{ab}(\theta-\theta_2-2(n-1)\pi)=-V_{ab}(\theta_2-\theta+2(n-1)\pi) $ & $(-1)^n\f{(t-l)^2}{128\pi^3t^3\ep^5}\left[(l+2t){\bf 1}_{ab}+\f{3(t+l)^2}{4t}\left(\gamma^1\gamma^t\right)_{ab}\right]$  \\ \hline
$V_{ab}(\theta-\theta_2+2\pi)=-V_{ab}( \theta_2-\theta-2\pi)$ & $\f{(t-l)^2}{128\pi^3t^3\ep^5}\left[-(l+2t){\bf 1}_{ab}-\f{3(t+l)^2}{4t}\left(\gamma^1\gamma^t\right)_{ab}\right]$  \\ \hline
    \end{tabular}}
  \end{center}
   \caption{The table of green function $S_{ab}(\theta_{\alpha}-\theta_{\beta})$ and $V_{ab}(\theta_{\alpha}-\theta_{\beta})$ for untwisted fields in $D=6$ dim..}
\end{table}

\begin{table}[h!]
  \begin{center} 
  \scalebox{0.8}{  
    \begin{tabular}{|c|c|} \hline
 $D=8,~~n=1$ & $$     \\ \hline
      $S_{ab}( \theta-\theta_2)=-S_{ab}(\theta_2-\theta) $ &  $-i\frac{3}{128 \pi ^4 \epsilon ^7}\gamma^t_{ab}$  \\ \hline
$V_{ab}( \theta-\theta_2)=-V_{ab}(\theta_2-\theta) $ &  $-\frac{3}{128 \pi ^4 \epsilon ^7}{\bf 1}_{ab}$  \\ \hline \hline
   
$n\ge2,~~t\le l$ &      \\ \hline
      $S_{ab}(\theta-\theta_2)=-S_{ab}(\theta_2-\theta) $ &  $-i\frac{3}{128 \pi ^4 \epsilon ^7}\gamma^t_{ab}$   \\ \hline
 $V_{ab}(\theta-\theta_2)=-V_{ab}(\theta_2-\theta) $ & $ -\frac{3}{128 \pi ^4 \epsilon ^7}{\bf 1}_{ab}$  \\ \hline \hline
  $t > l$      &    \\ \hline
     $S_{ab}(\theta-\theta_2)=-S_{ab}(\theta_2-\theta)$ & $-i\frac{3 (l+t)^3 \left(3 l^2-9 l t+8 t^2\right)}{2048 \pi ^4 t^5 \epsilon ^7}\gamma^t_{ab}+i\frac{15 \left(l^2-t^2\right)^3}{4096 \pi ^4 t^6 \epsilon ^7}\gamma^1_{ab}$  \\ \hline
$S_{ab}(\theta-\theta_2-2(n-1)\pi)=-S_{ab}(\theta_2-\theta+2(n-1)\pi) $ & $(-1)^{n-1}i\frac{3 (l-t)^3 \left(3 l^2+9 l t+8 t^2\right)}{2048 \pi ^4 t^5 \epsilon ^7}\gamma^t_{ab}-(-1)^{n-1}i\frac{15 \left(l^2-t^2\right)^3}{4096 \pi ^4 t^6 \epsilon ^7}\gamma^1_{ab}$  \\ \hline
$S_{ab}(\theta-\theta_2+2\pi)=-S_{ab}( \theta_2-\theta-2\pi)$ & $i\frac{3 (l-t)^3 \left(3 l^2+9 l t+8 t^2\right)}{2048 \pi ^4 t^5 \epsilon ^7}\gamma^t_{ab}+i\frac{15 \left(t^2-l^2\right)^3}{4096 \pi ^4 t^6 \epsilon ^7}\gamma^1_{ab}$  \\ \hline \hline
     $V_{ab}(\theta-\theta_2)=-V_{ab}(\theta_2-\theta)$ & $-\frac{3 (l+t)^3 \left(3 l^2-9 l t+8 t^2\right)}{2048 \pi ^4 t^5 \epsilon ^7}{\bf 1}_{ab}-\frac{15 \left(l^2-t^2\right)^3}{4096 \pi ^4 t^6 \epsilon ^7}\left(\gamma^1\g^t\right)_{ab}$  \\ \hline
$V_{ab}(\theta-\theta_2-2(n-1)\pi)=-V_{ab}(\theta_2-\theta+2(n-1)\pi) $ & $(-1)^{n-1}\frac{3 (l-t)^3 \left(3 l^2+9 l t+8 t^2\right)}{2048 \pi ^4 t^5 \epsilon ^7}{\bf 1}_{ab}+(-1)^{n-1}\frac{15 \left(l^2-t^2\right)^3}{4096 \pi ^4 t^6 \epsilon ^7}\left(\g^1\g^t\right)_{ab}$  \\ \hline
$V_{ab}(\theta-\theta_2+2\pi)=-V_{ab}( \theta_2-\theta-2\pi)$ & $\frac{3 (l-t)^3 \left(3 l^2+9 l t+8 t^2\right)}{2048 \pi ^4 t^5 \epsilon ^7}{\bf 1}_{ab}-\frac{15 \left(t^2-l^2\right)^3}{4096 \pi ^4 t^6 \epsilon ^7}\left(\g^1\g^t\right)_{ab}$  \\ \hline \hline
    \end{tabular}}
  \end{center}
   \caption{The table of green function $S_{ab}(\theta_{\alpha}-\theta_{\beta})$ and $V_{ab}(\theta_{\alpha}-\theta_{\beta})$ for untwisted fields in $D=8$ dim..}
\end{table}

\section{2d CFT details}\label{App:2d}
To compute above correlation function in section 4, we need the six types of correlation functions 
\be
\f{\left \langle e^{i\phi}(z_1) e^{- i \phi}(z_2) e^{i \phi} (z_3) e^{-i \phi}(z_4) e^{-i  2\alpha \phi} (z_5)e^{i 2\alpha \phi}(z_6) \right\rangle}{\left \langle e^{-2 i \alpha \phi}(z_5) e^{ 2i \alpha \phi}(z_6)\right\rangle} = \f{16 |z_1|^2 |z_2|^2}{|z_{12}|^4 |z_1 + z_2| ^4} \Big | \f{z_1 }{z_2} \Big| ^{8 \alpha}
\ee
\be
\f{\left\langle e^{-i\phi}(z_1) e^{i \phi}(z_2) e^{-i \phi} (z_3) e^{i \phi}(z_4) e^{-2i  \alpha \phi} (z_5)e^{2i \alpha \phi}(z_6) \right\rangle}{\left \langle e^{- 2i \alpha \phi}(z_5) e^{2i \alpha \phi}(z_6)\right\rangle} = \f{16 |z_1|^2 |z_2|^2}{|z_{12}|^4 |z_1 + z_2| ^4} \Big | \f{z_2 }{z_1} \Big| ^{8 \alpha}
\ee
\be
\f{\left\langle e^{i\phi}(z_1) e^{-i \phi}(z_2) e^{-i \phi} (z_3) e^{i \phi}(z_4) e^{-2i  \alpha \phi} (z_5)e^{2i \alpha \phi}(z_6) \right\rangle}{\left \langle e^{- 2i \alpha \phi}(z_5) e^{2i \alpha \phi}(z_6)\right\rangle} = \f{ |z_1+z_2|^4}{16|z_{12}|^4 |z_1|^2 | z_2| ^2} 
\ee
\be
\f{\left \langle e^{-i\phi}(z_1) e^{i \phi}(z_2) e^{i \phi} (z_3) e^{-i \phi}(z_4) e^{-2i  \alpha \phi} (z_5)e^{2i \alpha \phi}(z_6) \right\rangle}{\left \langle e^{- 2i \alpha \phi}(z_5) e^{2i \alpha \phi}(z_6)\right\rangle} = \f{ |z_1+z_2|^4}{16|z_{12}|^4 |z_1|^2 | z_2| ^2} 
\ee
\be
\f{\left \langle e^{i\phi}(z_1) e^{i \phi}(z_2) e^{-i \phi} (z_3) e^{-i \phi}(z_4) e^{-2i  \alpha \phi} (z_5)e^{2i \alpha \phi}(z_6) \right\rangle}{\left \langle e^{-2 i \alpha \phi}(z_5) e^{2i \alpha \phi}(z_6)\right\rangle} = \f{ |z_{12}|^4}{16|z_1+z_2|^4 |z_1|^2 | z_2| ^2} 
\ee
\be
\f{\left\langle e^{-i\phi}(z_1) e^{-i \phi}(z_2) e^{i \phi} (z_3) e^{i \phi}(z_4) e^{-2i  \alpha \phi} (z_5)e^{2i \alpha \phi}(z_6) \right\rangle}{\left \langle e^{- 2i \alpha \phi}(z_5) e^{2i \alpha \phi}(z_6)\right\rangle} = \f{ |z_{12}|^4}{16|z_1+z_2|^4 |z_1|^2 | z_2| ^2} 
\ee
where we take the limit of $z_5 \to 0$ and $z_6 \to \infty$.

\section{$\Delta S^{(n), c}_A$ for $\alpha_E$} \label{Appim}
Here we summarize the results for $\Delta S^{(n), c}_A$ with pure imaginary chemical potential $\alpha_E$.
\subsection*{Time evolution of $S^{(n), c}_A$}
The time dependence of $\Delta S^{(n), c}_A$ with pure imaginary chemical potential $\alpha_E$ for various states is summarized in Table 15 for three different fermionic operators. The results can be written in the form
\be
\Delta S^{(n), c}_A=\f{1}{1-n}\log{\left[A_1+A_2\right]}
\ee
and the explicit expressions for the three examples are in Table 15.
\begin{table}[htb]
  \begin{center} 
    \scalebox{0.8}{\begin{tabular}{|c|c|c|} \hline
      $\mathcal{O}$ & $\Delta S^{(n), c}_A$&The time value of $\Delta S^{(n), c}_A~~(t \rightarrow \infty)$     \\ \hline \hline
      $\psi_a$ & $\f{1}{1-n}\log{\left[A_1+A_2\right]}$ & $\f{1}{1-n}\log{\left[L_1+L_2\right]}$ \\ \hline
                  & $A_1=\f{(t+l)^n\left[2-\frac{l-t}{t}\left(\g^t \g^1\right)_{aa}\right]^n}{\left[2\left\{\left(t-l\right)e^{2i\pi \alpha_E}+(t+l)\right\}-\frac{(-l+t)(l+t)}{t}\left(e^{2i\pi \alpha_E}-1\right)\left(\g^t \g^1\right)_{aa}\right]^n}$ & $L_1=\f{\left[2+\left(\g^t\g^1\right)_{aa}\right]^n}{\left[2\left(e^{i 2 \pi \alpha_E}+1\right)-\left(e^{i 2 \pi \alpha_E}-1\right)\left(\g^t\g^1\right)_{aa}\right]^n}$ \\ \hline
& $A_2=\f{(t-l)^n\left[2-\frac{l+t}{t}\left(\g^t \g^1\right)_{aa}\right]^ne^{2ni \pi \alpha_E}}{\left[2\left\{\left(t-l\right)e^{2i\pi \alpha_E}+(t+l)\right\}-\frac{(-l+t)(l+t)}{t}\left(e^{2i\pi \alpha_E}-1\right)\left(\g^t \g^1\right)_{aa}\right]^n}$ & $L_2=\f{\left[2-\left(\g^t\g^1\right)_{aa}\right]^ne^{2ni \pi \alpha_E}}{\left[2\left(e^{i 2 \pi \alpha_E}+1\right)-\left(e^{i 2 \pi \alpha_E}-1\right)\left(\g^t\g^1\right)_{aa}\right]^n}$ \\ \hline \hline
$\bar{\psi}\psi$ & $\f{1}{1-n}\log{\left[A_1+A_2\right]}$ & \\ \hline
& $A_1 = \f{(t+l)^{2n}\left[4^2-4\left(\f{t-l}{t}\right)^2\right]^n+(t-l)^{2n}\left[4^2-4\left(\f{t+l}{t}\right)^2\right]^n}{\left[4^2\left\{(t-l)^2+(t+l)^2+2(t^2-l^2)\cos{\left(2\pi \alpha_E\right)}\right\}+8\left(\f{t^2-l^2}{t}\right)^2(\cos{\left(2\pi \alpha_E\right)}-1)\right]^n}$ & $\f{1}{1-n}\log{\left[ \frac{(12)^n+(12)^n+8 \sum_{k \in 2{\bf Z}}^{n \ge k}{}_n C_k5^{n-k}4^k\cos{\left(2n\pi \alpha_E\right)}}{\left[4^2\left(2+2\cos{(2\pi \alpha_E)}\right)+8\left(\cos{\left(2\pi \alpha_E\right)}-1\right)\right]^n}\right]}$ \\ \hline
& $A_2=\f{8(t^2-l^2)^n\sum^{n\ge k}_{k \in 2{\bf Z}}{}_n C_k4^k\left(4+\f{t^2-l^2}{t^2}\right)^{n-k}\cos{\left(2n\pi \alpha_E\right)}}{\left[4^2\left\{(t-l)^2+(t+l)^2+2(t^2-l^2)\cos{\left(2\pi \alpha_E\right)}\right\}+8\left(\f{t^2-l^2}{t}\right)^2(\cos{2\pi \alpha_E}-1)\right]^n}$ &  \\ \hline \hline
$\psi^{\dagger}\psi$ & $\f{1}{1-n}\log{\left[A_1+A_2\right]}$ & $$ \\ \hline
 & $A_1 = \f{(t+l)^{2n}\left[4^2+4\left(\f{t-l}{t}\right)^2\right]^n+(t-l)^{2n}\left[4^2+4\left(\f{t+l}{t}\right)^2\right]^n}{\left[4^2\left\{(t-l)^2+(t+l)^2+2(t^2-l^2)\cos{\left(2\pi \alpha_E\right)}\right\}+8\left(\f{t^2-l^2}{t}\right)^2(-\cos{\left(2\pi \alpha_E\right)}+1)\right]^n},$ & $\f{1}{1-n}\log{\left[ \frac{(20)^n+(20)^n+8\cdot 3^n\cos{\left(2n\pi \alpha_E\right)}}{\left[4^2\left(2+2\cos{(2\pi \alpha_E)}\right)+8\left(-\cos{\left(2\pi \alpha_E\right)}+1\right)\right]^n}\right]}$ \\ \hline
& $A_2=\f{8(t^2-l^2)^n\sum^{n\le k}_{k \in 2{\bf Z}}{}_n C_k\left(4\f{l}{t}\right)^k\left(4+\f{-t^2+l^2}{t^2}\right)^{n-k}\cos{\left(2n\pi \alpha_E\right)}}{\left[4^2\left\{(t-l)^2+(t+l)^2+2(t^2-l^2)\cos{\left(2\pi \alpha_E\right)}\right\}+8\left(\f{t^2-l^2}{t}\right)^2(-\cos{(2\pi \alpha_E)}+1)\right]^n}$ & \\ \hline
    \end{tabular}}
  \end{center}
   \caption{The table of $\Delta S^{(n), c}_A$ with $\alpha_E$.}
\end{table}

\subsection*{Physical interpretation}
In this subsection, we reproduce the results in the previous subsection under quasi-particle interpretation.
We decompose $\psi_a, \psi_a^{\dagger}$ and $\bar{\psi}_a$ into right and left movers respectively as in (\ref{DQ}).

In $4d$, anti-commutation relationships for left and right movers are given by (\ref{DQ}). Therefore the reduced density matrices $\rho_A^p, \rho_A^{bp}$ and $\rho^{dp}_A$ for $\psi_a, \bar{\psi}\psi$ and $\psi^{\dagger}\psi$ are respectively given by Table 6.

The charged reduced density matrix $\rho_A^c$ with pure imaginary entanglement chemical potential $\alpha_E$ is define by 
\be
\begin{split}
\rho_A^c = \f{e^{-i2\pi \alpha_E Q_A}\rho_A}{\Tr_A e^{-i2\pi \alpha_E Q_A}\rho_A}.
\end{split}
\ee
where its anti-commutation relationships are given by
\be
\begin{split}
[Q_A, \psi_a]=- \psi^R_a,~~~~[Q_A, \bar{\psi}^{R, \dagger}_a]=\bar{\psi}^{R, \dagger}_a,~~~~[Q_A, \phi^{R, \dagger}_a]=\phi^{R, \dagger}_a
\end{split}
\ee 

Therefore the reduced density matrices  $\rho_A^p, \rho_A^{bp}$ and $\rho^{dp}_A$ are given by Table 6. $C, D$ and $E$ are listed up in Table 16.
\begin{table}[htb]
  \begin{center}
    \begin{tabular}{|c|c|c|c|} \hline
      $C_{0, R}$ & $\f{\left(2+\left(\g^t\g^1\right)_{aa}\right)}{\left[2(1+e^{2i\pi \alpha_E})+(1-e^{2i\pi \alpha_E})\left(\g^t\g^1\right)_{aa}\right]}$ &$C_{1, R}$ &$\f{e^{2i \pi \alpha_E}\left(2-\left(\g^t\g^1\right)_{aa}\right)}{\left[2(1+e^{2i\pi \alpha_E})+(1-e^{2i\pi \alpha_E})\left(\g^t\g^1\right)_{aa}\right]}$ \\ \hline \hline
   $D_{0, R}=D_{3, R}$ & $\f{12}{24+40 \cos{\left(2\pi \alpha_E\right)}}$ &$D_{1, R, 1}=D_{1, R, 3}$ &$\f{9 e^{2i\pi \alpha_E}}{24+40 \cos{\left(2\pi \alpha_E\right)}}$ \\ \hline
$D_{1, R, 2}=D_{1, R, 4}$ & $\f{ e^{2i\pi \alpha_E}}{24+40 \cos{\left(2\pi \alpha_E\right)}}$ &$D_{2, R, 1}=D_{2, R, 3}$& $ \f{9 e^{-2i\pi \alpha_E}}{24+40 \cos{\left(2\pi \alpha_E\right)}}$ \\ \hline 
$D_{2, R, 2}=D_{2, R, 4}$ & $\f{ e^{-2i\pi \alpha_E}}{24+40 \cos{\left(2\pi \alpha_E\right)}}$ & & \\ \hline \hline
$E_{0, R}=E_{1, R}$ & $\f{20}{40+24\cos{(2\pi \alpha_E)}}$ &$E_{2, R}$ &$\f{3e^{i2 \pi \alpha_E}}{40+24\cos{(2\pi \alpha_E)}}$ \\ \hline
$E_{3, R}$ & $\f{3e^{-i2 \pi \alpha_E}}{40+24\cos{(2\pi \alpha_E)}}$ & & \\ \hline
    \end{tabular}
  \end{center}
  \caption{The table of $C, D$ and $E$.}
\end{table}

By using the reduced density matrices in Table 6 and 16, $\Delta S^{(n), c}_A$ are consistent with those in the previous section.

$\Delta S^{(n), c}_A$ for $\rho_A^{p, c}, \rho_A^{bp, c}, \rho_A^{dp, c}$ are respectively listed up  
Table 17.
\begin{table}[htb]
  \begin{center}
    \begin{tabular}{|c|c|} \hline
      $\mathcal{O}$ & $\Delta S^{(n), c}_A$ \\ \hline \hline
   $\psi_a$ & $\frac{1}{1-n}\log \left(\frac{\left(2-\left(\g^t\g^1\right)_{aa}\right)^n e^{i2n\pi \alpha_E  }+\left(\left(\g^t\g^1\right)_{aa}+2\right)^n}{\left(\left(\g^t\g^1\right) \left(1-e^{i2\pi \alpha_E}\right)+2 \left(e^{i2\pi \alpha_E}+1\right)\right)^n}\right)$  \\ \hline 
$\bar{\psi}\psi$ & $\frac{1}{1-n}\log \left(\frac{4\cdot 9^n \cos (2 \pi  \alpha_E  n)+4 \cos (2 \pi  \alpha_E  n)+2\ 12^n}{(40 \cos (2 \pi  \alpha_E )+24)^n}\right)$  \\ \hline 
$\psi^{\dagger}\psi$ & $ \frac{1}{1-n}\log \left(\frac{8\ 3^n \cos (2 \pi  \alpha_E  n)+2\ 20^n}{(24 \cos (2 \pi  \alpha_E )+40)^n}\right)$  \\ \hline 
    \end{tabular}
  \end{center}
   \caption{$\Delta S^{(n), c}_A$ with $\alpha_E$ for $\rho_A^{p, c}, \rho_A^{bp, c}, \rho_A^{dp, c}$.}
\end{table}
 If we take the von Neumann limit $n \rightarrow 1$, $\Delta S^{c}_A$ for $\psi_a, \bar{\psi}\psi$ and $\psi^{\dagger} \psi$ are respectively given by
\be
\begin{split}
&\Delta S^{p, c}_A =\f{2+\left(\g^t\g^1\right)_{aa}}{F_1}\log{\left[\f{F_1}{2+\left(\g^t\g^1\right)_{aa}}\right]}+\f{\left(2-\left(\g^t\g^1\right)_{aa}\right)e^{i2\pi\alpha_E}}{F_1}\log{\left[\f{F_1}{\left(2-\left(\g^t\g^1\right)_{aa}\right)e^{i2\pi\alpha_E}}\right]}, \\
&\Delta S^{bp, c}_A =\f{3}{F_2}\log{\left[\f{2F_2}{3}\right]}+\f{5\cos{\left(2\pi\alpha_E\right)}}{F_2}\log{\left[8F_2\right]}-\f{9\cos{2\pi\alpha_E}}{F_2}\log{3} \\
&~~~~~~~~~~~~~~~~~~~~~~~~~~~~~~~~~~~~~~~~~~~~~~~~~~~~~~~~+\f{5(e^{-i2\pi\alpha_E}\log{e^{i2\pi\alpha_E}}+e^{i2\pi\alpha_E}\log{e^{-i2\pi\alpha_E}})}{2F_2}, \\
&\Delta S^{dp, c}_A=\f{5}{F_3}\log{\left[\f{2F_3}{5}\right]}+\f{3\cos{\left(2\pi\alpha_E\right)}}{F_3}\log{\left[\f{8F_3}{3}\right]}+\f{3(e^{-i2\pi\alpha_E}\log{e^{i2\pi\alpha_E}}+e^{i2\pi\alpha_E}\log{e^{-i2\pi\alpha_E}})}{2F_3}, \\
\end{split}
\ee
where the functions $F_i$ are given by
\be
\begin{split}
&F_1=2+\left(\g^t\g^1\right)_{aa}+e^{i2\pi\alpha_E}(2-\left(\g^t\g^1\right)_{aa}),\\
&F_2=3+5\cos{2\pi\alpha_E}, \\
&F_3=5+3\cos{2\pi\alpha_E}.
\end{split}
\ee
$\Delta S^{p, c}_A, \Delta S^{bp, c}_A$ and $\Delta S^{dp, c}_A$ respectively correspond to $\psi_a, \bar{\psi}\psi$ and $\psi^{\dagger}\psi$.

\subsection*{$\Delta S^{(n), c}_A$ with $\alpha_E$ in $D$-dimensional Spacetime in Large $N$ limit}
Here we consider $\Delta S^{(n)}_A$ for $(\psi_a)^l_k, \Tr\left(\bar{\psi}\psi\right)$ and $\Tr\left(\psi ^{\dagger} \psi\right)$.
In $D$ dim. $U(N)$ free massless fermionic field theory, fermionic fields are decomposed into the left and right moving modes as in (\ref{DQ}) and anti-commutation relation (\ref{gdac}) for them are imposed.
Let's derive $\Delta S^{(n), c}_A$ for $\rho^{p, c}_A, \rho^{bp, c}_A$ and $\rho^{dp, c}_A$.
$\Delta S^{(n), c}_A$ for $\left(\psi_a\right)^{i}_j$, $\Tr\bar{\psi}\psi$ and $\Tr\psi^{\dagger}\psi$ are listed up in Table 18.

\begin{table}[h!]
  \begin{center}
\scalebox{0.7}{\begin{tabular}{|c|c|c|} \hline
      $\mathcal{O}$ &$\Delta S^{(n), c}_A$ for $D\neq2$  &$\Delta S^{(n), c}_A$ for $D=2$, $\left(\g^t\g^1\right)_{aa}\neq \pm 1$\\ \hline \hline
   $\psi_a$ & $\f{1}{1-n}\log{\left[\f{\left(1+c_g \left(\g^t \g^1\right)_{aa}\right)^n+e^{i 2 \pi \alpha n}\left(1-c_g \left(\g^t \g^1\right)_{aa}\right)^n}{1+e^{i2\pi \alpha}+c_g (1- e^{i2 \pi \alpha})\left(\g^t \g^1\right)_{aa}}\right]}$ &$\f{1}{1-n}\log{\left[\f{\left(1+ \left(\g^t \g^1\right)_{aa}\right)^n+e^{i 2 \pi \alpha n}\left(1- \left(\g^t \g^1\right)_{aa}\right)^n}{1+e^{i2\pi \alpha}+ (1- e^{i2 \pi \alpha})\left(\g^t \g^1\right)_{aa}}\right]}$  \\ \hline 
$\Tr\bar{\psi}\psi$ & $\f{1}{1-n}\log{\left[\f{2\cdot 2^{\f{D n}{2}}(1-c_g^2)^nN^{2n}+2^{\f{D}{2}}N^2\cos{\left(2n \pi \alpha\right)\left\{(1-c_g)^{2n}+(1+c_g)^{2n}\right\}}}{2^{\f{(D+2)n}{2}}N^{2n}\left\{\left(1-c_g^2\right)+(1+c_g^2)\cos{\left(2\pi \alpha\right)}\right\}^n}\right]}$ &$\log{2}+2\log{N}+\f{1}{1-n}\log{\left[\f{\cos{2n\pi\alpha}}{(\cos{2\pi\alpha})^n}\right]}$ \\ \hline 
$\Tr\psi^{\dagger}\psi$ & $ \f{1}{1-n}\log{\left[\frac{2\cdot 2^{\f{n D}{2}}(1+c_g^2)^n N^{2n}+2\cdot2^{\f{D}{2}}(1-c_g^2)^n\cos{\left(2n\pi\alpha\right)}N^2}{2^{\f{D+2}{2}n}\left[(1+c_g^2)+(1-c_g^2)\cos{\left(2\pi \alpha\right)}\right]^n N^{2n}}\right]}$ & $\log{2}$ \\ \hline \hline
 $\mathcal{O}$ & $\Delta S^{c}_A$ for $D \neq2$ & $\Delta S^{c}_A$ for $D =2$   \\ \hline
$\left(\psi_a\right)^{i}_j$ & $\f{1+c_g\left(\g^t\g^1\right)_{aa}}{F_1}\log{\left(\f{F_1}{1+c_g\left(\g^t\g^1\right)_{aa}}\right)}+\f{\left(1-c_g\left(\g^t\g^1\right)_{aa}\right)}{F_1e^{-i2\pi \alpha_E}}\log{\left(\f{F_1e^{-2i\pi \alpha_E}}{\left(1-c_g\left(\g^t\g^1\right)_{aa}\right)}\right)}$ & $\f{1+\left(\g^t\g^1\right)_{aa}}{F_4}\log{\left(\f{F_4}{1+\left(\g^t\g^1\right)_{aa}}\right)}+\f{\left(1-\left(\g^t\g^1\right)_{aa}\right)}{F_4e^{-2i\pi \alpha_E}}\log{\left(\f{F_4e^{-2i\pi \alpha_E}}{\left(1-\left(\g^t\g^1\right)_{aa}\right)}\right)}$\\
 & where $F_1=1+c_g\left(\g^t\g^1\right)_{aa}+(1-c_g\left(\g^t\g^1\right)_{aa})e^{i2\pi \alpha_E}$ & $F_4=1+\left(\g^t\g^1\right)_{aa}+(1-\left(\g^t\g^1\right)_{aa})e^{2i\pi \alpha_E}$\\ \hline
$\Tr\bar{\psi}\psi$  & $\f{(1-c_g^2)}{F_2}\log{\left(\f{2F_2}{(1-c_g^2)}\right)}+\f{(1-c_g)^2\cos{(2\pi \alpha_E)}}{2F_2}\log{\left(\f{2^{\f{D+2}{2}}N^2F_2}{(1-c_g)^2}\right)}$ & $$ \\ 
& $+\f{(1+c_g)^2\cos{(2\pi \alpha_E)}}{2F_2}\log{\left(\f{2^{\f{D+2}{2}}N^2F_2}{(1+c_g)^2}\right)}$ & $\f{1}{2\cos{\left(2\pi\alpha_E\right)}}\left[e^{i2\pi \alpha_E}\log{e^{-i2\pi \alpha_E}}+e^{-i2\pi \alpha_E}\log{e^{i2\pi \alpha_E}}\right]$ \\
& $+\f{ (1+c_g^2)}{2F_2}\left[e^{i2\pi \alpha_E}\log{e^{-i2\pi \alpha_E}}+e^{-i2\pi \alpha_E}\log{e^{i2\pi \alpha_E}}\right]$ & $+\log{\left(2N^2\cos{2\pi \alpha_E}\right)}$  \\
& where $F_2=(1-c_g^2)+(1+c_g^2)\cos{2\pi \alpha_E}$ & $$ \\ \hline
$\Tr\psi^{\dagger}\psi$   & $\f{1+c_g^2}{F_3}\log{\left(\f{2F_3}{1+c_g^2}\right)}+\f{(1-c_g^2)\cos{(2\pi \alpha_E)}}{F_3}\log{\left( \f{N^22^{\f{D+2}{2}}F_3}{(1-c_g^2)}\right)}$ & $$  \\
   &~~~~~~~~~~~~~~~~~~~~~~~~~~~~~~$+\f{ (1-c_g^2)}{2F_3}\left[e^{i2\pi \alpha_E}\log{e^{-i2\pi \alpha_E}}+e^{-i2\pi \alpha_E}\log{e^{i2\pi \alpha_E}}\right]$  & $\log{2}$ \\
 & where $F_3=1+c_g^2+(1-c_g^2)\cos{2\pi \alpha_E}$ & $$ \\ \hline
    \end{tabular}}
  \end{center}
  \caption{$\Delta S^{(n), c}_A$ with $\alpha_E$ for $\rho_A^{p, c}, \rho_A^{bp, c}, \rho_A^{dp, c}$ in $D$ dimensional spacetime. }
\end{table}

\subsubsection*{Large $N$ Limit}

Here we study $\Delta S^{(n), c}_A$ with the imaginary chemical potential for $\left(\psi_a\right)^i_j, \Tr\bar{\psi}\psi$ and $\Tr\psi^{\dagger}\psi$.
$\Delta S^{(n), c}_A$ does not depend on $N$.
If we take large $N$ limit for $\cos{2\pi \alpha_E}\neq 0$, $\Delta S^{(n), c}_A$ for $\Tr\bar{\psi}\psi$ in $D \neq 2$ reduce to
\be
\begin{split}
&\Delta S^{(n\ge2), c}_A\sim\f{1}{1-n}\log{\left[\f{2\left(1-c_g^2\right)^n}{2^n\left\{(1-c_g^2)+(1+c_g^2)\cos{\left(2\pi \alpha_E\right)}\right\}^n}\right]}, \\
&\Delta S^{c}_A \sim\frac{2\cos (2 \pi  \alpha ) \left(\left(c_g^2+1\right)\right)}{ \left(\left(c_g^2+1\right) \cos (2 \pi  \alpha_E )-c_g^2+1\right)}  \log (N).
\end{split}
\ee
If we take large $N$ limit for $\cos{2\pi \alpha_E}\neq 0$ and $\cos{2n\pi \alpha_E}\neq 0$, $\Delta S^{(n), c}_A$ for $\Tr\bar{\psi}\psi$ in $D=2$ are given by
\be
\Delta S^{(n), c}_A\sim 2\log{N}
\ee
where $n$ is arbitrary integers.
$\Delta S^{(n), c}_A$ are (R$\acute{e}$nyi) entanglement entropies for a maximally entangled state.
It is for the $N^2 \times N^2$ reduced density matrix. 

If the large $N$ limit is taken for $\cos{2\pi \alpha_E} \neq 0$, $\Delta S^{(n)}_A$ for $\Tr\psi^{\dagger}\psi$ in $D\neq2$ reduce to
\be
\begin{split}
&\Delta S^{(n \ge 2), c}_A \sim\f{1}{1-n}\log{\left[\f{2(1+c_g^2)^n}{\left[2\left\{(1+c_g^2)+(1-c_g^2)\cos{\left(2\pi \alpha\right)}\right\}\right]^n}\right]}, \\
&\Delta S^{c}_A\sim\frac{2\left(c_g^2-1\right) \cos (2 \pi  \alpha ) }{ \left(c_g^2-1\right) \cos (2 \pi  \alpha )- \left(c_g^2+1\right)} \log (N). \\
\end{split}
\ee
Even in the large $N$ limit, entropies for $\Tr\psi^{\dagger}\psi$ in $D=2$ dim. are given by
\be
\Delta S^{(n)}_A = \log{2}
\ee
which does not depend on $\alpha_E$, $N$ and $n$ and correspond to the answer for the EPR state.

\section{Example in $AdS_3/CFT_2$}
In this appendix we present another example of R\'enyi entropy with charge that can be treated analytically. In \cite{m3} and in the main text above we used holography to compute the 2n-point correlator on $\Sigma_n$ in a CFT at large central charge. More precisely, we used the geodesic approximation in the background of a topological black hole that after identification of the temperature $\beta=2\pi n$ has $\Sigma_n$ at the boundary. In three dimensions, formally, there are no topological black holes but we can simply take the BTZ black hole with the above identification of the inverse temperature and reproduce the large $c$ result on $\Sigma_n$.\\
Here we generalize this to the spinning BTZ black hole\footnote{In $AdS_3/CFT_2$ context spinning BTZ black holes usually appear with a constant CS gauge field in the bulk and are dual to a 2d CFT with additional U(1) Kac-Moody current (see \cite{Kraus:2006wn} for review and \cite{Caputa:2013eka} for results on evolution of entanglement).}. The spin in the dual geometry introduces left-right asymmetry in the CFT that lives on a twisted cylinder (we will consider a BTZ black string). After we identify the Hawking temperature appropriately we can again use holography to compute the 2n-point correlator on this twisted $\Sigma_n$. This computation turns out to be quite subtle and we only present a short summary below and we hope to come back to it elsewhere \cite{wip} with natural generalization to higher spin black holes.

\medskip
Let us start with Lorentzian black hole \cite{Banados:1992wn} 
\be
ds^2=-\frac{(r^2-r^2_+)(r^2-r^2_-)}{r^2}dt^2+
\frac{r^2}{(r^2-r^2_+)(r^2-r^2_-)}dr^2+r^2
\left(d\phi -\frac{r_+r_-}{r^2}dt\right)^2,
\label{LBTZJ}
\ee
where the mass $M$, the angular momentum $J$, the Hawking temperature $T_H=1/\beta_H$ and the angular velocity $\Omega$ of the outer horizon are determined by radii of the inner ($r_-$) and outer ($r_+$) horizons, as follows
\be
M=r^2_++r^2_-,\quad J=2\,r_+r_-, \quad 
\beta_H=1/T_H = \frac{2\pi r_+}{r_+^2 - r_-^2}, 
\quad \Omega = \frac{r_-}{r_+},
\ee
one can also define left and right temperatures via $r_{\pm}=\pi\left(T_+\pm T_-\right)$.\\
The Euclidean black hole is obtained by analytic continuation to imaginary time and imaginary angular momentum (usual momentum for the black string)
\be
t\to i\tau_E,\qquad J\to i J_E,
\ee
such that we have
\be 
ds^2_E=\frac{(r^2-r^2_+)(r^2-r^2_-)}{r^2}d\tau^2_E+
\frac{r^2}{(r^2-r^2_+)(r^2-r^2_-)}dr^2+r^2
\left(d\phi -\frac{r_+(ir_-)}{r^2}d\tau_E\right)^2.
\label{BTZJ}
\ee 
For real $\tau_E$ and $J_E$, the quantities $ds^2_E$, $r^2_{\pm}$ and $r_+$ are real and $r_-$ is pure imaginary. We then have $u=u_+=\phi+i\tau_E$ and $\bar{u}=u_-=\phi-i\tau_E$ that are complex conjugates of each other and similarly $T_+=T$ and $T_-=\bar{T}$. The radial coordinate ranges between $r_+\le r\le \infty$, and we have the identification
\bea
&&(u,r)\simeq (u+2\pi,r)\simeq(u+i\beta,r) \nn\\
&&(\bar{u},r)\simeq (\bar{u}+2\pi,r)\simeq(u-i\bar{\beta},r) ,
\eea
where $\beta=1/T$ and $\bar{\beta}=1/\bar{T}$ are complex. The boundary of the Euclidean BH is a torus with modular parameter $\frac{i\beta}{2\pi}$. Note also that since $r_-$ is pure imaginary the the potential $\Omega=i\Omega_E$ is also purely imaginary. In the following we will consider the BTZ black string which has the same metric but with $\phi$ decompactified (no $2\pi$ identification on $u$). \\
Now we generalize the computation of the correlation functions using the topological black hole. Namely we identify $\beta_H=2\pi n$ and ''naively" compute the correlator holographically using geodesic approximation. For that, we assume that all the analytic continuations as well as the appropriate ranges of $\Omega$ are taken.\\
The length of the geodesic between points separated by $\Delta\phi$ and $\Delta\tau$ at radial cutoff $\epsilon_{uv}$ can be written as \cite{Hemming:2002kd}
\be
\frac{L}{R}=\log\left(\frac{\beta^2_H(1-\Omega^2)}{\pi^2\epsilon^2_{uv}}\sinh\frac{\pi(\Delta\phi+i\Delta\tau)}{\beta_H(1-\Omega)}\sinh\frac{\pi(\Delta\phi-i\Delta\tau)}{\beta_H(1+\Omega)}\right).
\ee
Therefore the relevant ratio of the propagators at large $c$ becomes
\be
e^{-\frac{2\Delta_O}{R}\left(L^{(n)}-L^{(1)}\right)}=\left[\frac{\sinh\frac{\Delta\phi+i\Delta\tau}{2(1-\Omega)}\sinh\frac{\Delta\phi-i\Delta\tau}{2(1+\Omega)}}{n^2\sinh\frac{\Delta\phi+i\Delta\tau}{2n(1-\Omega)}\sinh\frac{\Delta\phi-i\Delta\tau}{2n(1+\Omega)}}\right]^{2\Delta_O}\label{ratio2}.
\ee
For $\Omega\to0$ this reduces to the result in \cite{m3}. This result can be reproduced from CFT by first considering the ratio of the two-point function on $\Sigma_n$ to the two-point function on $\Sigma_1$
\be
\frac{\langle O(w_1,\bar{w}_1)O(w_2,\bar{w}_2)\rangle_{\Sigma_n}}{\langle O(w_1,\bar{w}_1)O(w_2,\bar{w}_2)\rangle_{\Sigma_1}}=\left|\frac{w_{12}}{w^{\frac{1}{2}}_1w^{\frac{1}{2}}_2}\right|^{4\Delta_O}\left|n\frac{w^{\frac{1}{n}}_1-w^{\frac{1}{n}}_2}{w^{\frac{1}{2n}}_1w^{\frac{1}{2n}}_2}\right|^{-4\Delta_O},
\ee
and applying the map 
\be
w=e^{\frac{\phi+i\tau}{1-\Omega}},\qquad \bar{w}=e^{\frac{\phi-i\tau}{1+\Omega}},
\ee
where in all the above formulas $\Omega$ is purely imaginary.\\
Now we determine which geodesics are dominant in the late time limit and non-zero $\Omega$. Generically, as in \cite{m3}, we first assume non-zero correlators between operators $\mathcal{O}$ and $\mathcal{O}^\dagger$. In this case, the leading contribution to the correlator comes from contractions are between operators on the same sheet\footnote{the contraction between the operator on the n-th and the 1st sheet becomes sub-leading due to the presence of $\Omega$}. Inserting the late time values \eqref{Ltime} and expanding to order $\epsilon$ gives
\be
e^{-\frac{2\Delta_O}{R}\left(L^{(n)}-L^{(1)}\right)}\simeq\left[\frac{\sin\frac{\pi}{1-\Omega}}{n\sin\frac{\pi}{n(1-\Omega)}}-\frac{\frac{\cos\frac{\pi}{1-\Omega}}{\cos\frac{\pi}{n(1-\Omega)}}-\frac{\sin\frac{\pi}{1-\Omega}}{n\sin\frac{\pi}{n(1-\Omega)}}}{(1-\Omega)\tan\frac{\pi}{n(1-\Omega)}}\frac{\epsilon}{n\,t}\right]^{2\Delta_O}.
\ee
Notice that the propagator only depends on the combination of $1-\Omega$ what can be understood in terms of the asymmetry between the left and right movers on $\Sigma_n$. Moreover, as long as $0<\Omega<\frac{1}{2}$, we can neglect the $O(\epsilon)$ term and the ratio approaches a constant at late times. For $\Omega\ge \frac{1}{2}$ there constant term vanishes for some specific values of $\Omega$ and we have to include the $\epsilon/t$ terms that lead to the logarithmic growth with time. Even for $\Omega <\f{1}{2}$ the propagators for the not-nearest operators can be $\mathcal{O}(1)$ at certain $\Omega$. However in this paper we do not consider $\Delta S^{(n), c}_A$ in those cases.  Similar subtleties appear in the less generic situations with e.g. $\mathcal{O}=\mathcal{O}^\dagger$. More detailed analysis of these interesting behavior deserves further study and we hope to come back to it in the future.\\  
Finally we can evaluate the 2n-point correlator using the geodesic approximation and inserting the late time values we have the change in the n-th R\'enyi (even $n$) entropy 
\be
\Delta S^{(n)}_A\simeq \frac{\Delta_O}{1-n}\log\left[\left(\frac{\sin\frac{\pi}{1-\Omega}}{n\sin\frac{\pi}{n(1-\Omega)}}\right)^{2n}\right],
\ee
and for $\Omega\neq 0$ we can take the von-Neumann limit $n\to 1$
\be
\Delta S^{(1)}_A=2\Delta_O\left(1-\frac{\pi}{1-\Omega}\cot\frac{\pi}{1-\Omega}\right).
\ee
It is also worth stressing that analysis of the R\'enyi entropies also breaks down in the extremal limit $\Omega\to 1$ which might be an interesting problem to explore in the future also in the context or higher spin black holes.



\begin{thebibliography}{}
\bibitem{Srednicki:1993im}
  M.~Srednicki,
  ``Entropy and area,''
  Phys.\ Rev.\ Lett.\  {\bf 71} (1993) 666
  doi:10.1103/PhysRevLett.71.666
  [hep-th/9303048].

\bibitem{Kitaev:2005dm}
  A.~Kitaev and J.~Preskill,
  ``Topological entanglement entropy,''
  Phys.\ Rev.\ Lett.\  {\bf 96} (2006) 110404
  doi:10.1103/PhysRevLett.96.110404
  [hep-th/0510092].
  
 \bibitem{Bhattacharya:2012mi}
  J.~Bhattacharya, M.~Nozaki, T.~Takayanagi and T.~Ugajin,
  ``Thermodynamical Property of Entanglement Entropy for Excited States,''
  Phys.\ Rev.\ Lett.\  {\bf 110} (2013) 9,  091602
  [arXiv:1212.1164].
  T.~Faulkner, M.~Guica, T.~Hartman, R.~C.~Myers and M.~Van Raamsdonk,
  ``Gravitation from Entanglement in Holographic CFTs,''
  JHEP {\bf 1403} (2014) 051
  [arXiv:1312.7856 [hep-th]].

\bibitem{Holzhey:1994we}
  C.~Holzhey, F.~Larsen and F.~Wilczek,
  ``Geometric and renormalized entropy in conformal field theory,''
  Nucl.\ Phys.\ B {\bf 424} (1994) 443
  doi:10.1016/0550-3213(94)90402-2
  [hep-th/9403108].
  
  
\bibitem{Calabrese:2004eu} 
  P.~Calabrese and J.~L.~Cardy,
  ``Entanglement entropy and quantum field theory,''
  J.\ Stat.\ Mech.\  {\bf 0406}, P06002 (2004)
  doi:10.1088/1742-5468/2004/06/P06002
  [hep-th/0405152].


\bibitem{Maldacena:1997re}
  J.~M.~Maldacena,
  ``The Large N limit of superconformal field theories and supergravity,''
  Int.\ J.\ Theor.\ Phys.\  {\bf 38} (1999) 1113
   [Adv.\ Theor.\ Math.\ Phys.\  {\bf 2} (1998) 231]
  doi:10.1023/A:1026654312961
  [hep-th/9711200].

\bibitem{RT}
  S.~Ryu and T.~Takayanagi,
  ``Holographic derivation of entanglement entropy from AdS/CFT,''
  Phys.\ Rev.\ Lett.\  {\bf 96} (2006) 181602; S.~Ryu and T.~Takayanagi,
 ``Aspects of holographic entanglement entropy,''
  JHEP {\bf 0608} (2006) 045;
 V.~E.~Hubeny, M.~Rangamani and T.~Takayanagi, ``A Covariant
holographic entanglement entropy proposal,'' JHEP {\bf 0707} (2007)
062  [arXiv:0705.0016 [hep-th]];
T.~Nishioka, S.~Ryu and T.~Takayanagi,
 ``Holographic Entanglement Entropy: An Overview,''
  J.\ Phys.\ A  {\bf 42} (2009) 504008;
 T.~Takayanagi,
  ``Entanglement Entropy from a Holographic Viewpoint,''
  Class.\ Quant.\ Grav.\  {\bf 29} (2012) 153001  [arXiv:1204.2450 [gr-qc]].

 
\bibitem{Islametal} 
R. Islam, R. Ma, P. M. Preiss, M. E. Tai, A. Lukin, M. Rispoli and M. Greiner. "Measuring entanglement entropy through the interference of quantum many-body twins" - arXiv:1509.01160  
   
   
  \bibitem{cag}
P.~Calabrese and J.~L.~Cardy,
  ``Evolution of Entanglement Entropy in One-Dimensional Systems,''
  J.\ Stat.\ Mech.\  {\bf 04} (2005) P04010, cond-mat/0503393.

\bibitem{lquench} 
  M.~Nozaki, T.~Numasawa and T.~Takayanagi,
  ``Holographic Local Quenches and Entanglement Density,''
  JHEP {\bf 1305}, 080 (2013)
  [arXiv:1302.5703 [hep-th]]. 


\bibitem{m2} 
  M.~Nozaki, T.~Numasawa and T.~Takayanagi,
  ``Quantum Entanglement of Local Operators in Conformal Field Theories,''
  Phys.\ Rev.\ Lett.\  {\bf 112}, 111602 (2014)
  [arXiv:1401.0539 [hep-th]].

\bibitem{m1} 
M.~Nozaki,
 ``Notes on Quantum Entanglement of Local Operators,''
  JHEP {\bf 1410}, 147 (2014)
  doi:10.1007/JHEP10(2014)147
  [arXiv:1405.5875 [hep-th]].
  
  \bibitem{He:2014mwa}
  S.~He, T.~Numasawa, T.~Takayanagi and K.~Watanabe,
  ``Quantum Dimension as Entanglement Entropy in 2D CFTs,''
  arXiv:1403.0702 [hep-th].
  
     \bibitem{m3} 
  P.~Caputa, M.~Nozaki and T.~Takayanagi,
  ``Entanglement of local operators in large-N conformal field theories,''
  PTEP {\bf 2014}, 093B06 (2014)
  [arXiv:1405.5946 [hep-th]].

 
    \bibitem{m4} 
  M.~Nozaki, T.~Numasawa and S.~Matsuura,
  ``Quantum Entanglement of Fermionic Local Operators,''
  arXiv:1507.04352 [hep-th].
 

\bibitem{LOP}
  N.~Shiba,
  ``Entanglement Entropy of Disjoint Regions in Excited States : An Operator Method,''
  arXiv:1408.0637 [hep-th];
  W.~Z.~Guo and S.~He,
  ``R\'enyi entropy of locally excited states with thermal and boundary effect in 2D CFTs,''
  JHEP {\bf 1504}, 099 (2015)
  [arXiv:1501.00757 [hep-th]];
  P.~Caputa and A.~Veliz-Osorio,
  ``Entanglement constant for conformal families,''
  Phys.\ Rev.\ D {\bf 92} (2015) 6,  065010
  [arXiv:1507.00582 [hep-th]];
  B.~Chen, W.~Z.~Guo, S.~He and J.~q.~Wu,
  ``Entanglement Entropy for Descendent Local Operators in 2D CFTs,''
  JHEP {\bf 1510} (2015) 173
  [arXiv:1507.01157 [hep-th]];
  W.~Z.~Guo,
  ``Coherent state, local excitation in 2D conformal field theory,''
  arXiv:1510.07142 [hep-th];
  P.~Caputa, J.~Simon, A.~Stikonas and T.~Takayanagi,
  ``Quantum Entanglement of Localized Excited States at Finite Temperature,''
  JHEP {\bf 1501}, 102 (2015)
  [arXiv:1410.2287 [hep-th]];
  P.~Caputa, J.~Simon, A.~Stikonas, T.~Takayanagi and K.~Watanabe,
  ``Scrambling time from local perturbations of the eternal BTZ black hole,''
  JHEP {\bf 1508}, 011 (2015)
  [arXiv:1503.08161 [hep-th]];
  C.~T.~Asplund, A.~Bernamonti, F.~Galli and T.~Hartman,
  JHEP {\bf 1502} (2015) 171
  [arXiv:1410.1392 [hep-th]].
  
\bibitem{hchee} 
  A.~Belin, L.~Y.~Hung, A.~Maloney, S.~Matsuura, R.~C.~Myers and T.~Sierens,
  ``Holographic Charged Renyi Entropies,''
  JHEP {\bf 1312}, 059 (2013)
  doi:10.1007/JHEP12(2013)059
  [arXiv:1310.4180 [hep-th]].

\bibitem{dw1} 
  J.~S.~Dowker,
  ``Charged Renyi entropies for free scalar fields,''
  arXiv:1512.01135 [hep-th];
  J.~S.~Dowker,
  ``Conformal weights of charged Renyi entropy twist operators for free scalar fields,''
  arXiv:1508.02949 [hep-th];
  J.~S.~Dowker,
  ``Conformal weights of charged Renyi entropy twist operators for free Dirac fields in arbitrary dimensions,''
  arXiv:1510.08378 [hep-th].

\bibitem{mtu1} 
  A.~Belin, L.~Y.~Hung, A.~Maloney and S.~Matsuura,
  ``Charged Renyi entropies and holographic superconductors,''
  JHEP {\bf 1501}, 059 (2015)
  doi:10.1007/JHEP01(2015)059
  [arXiv:1407.5630 [hep-th]].

\bibitem{Line} 
  B.~Linet,
  ``Euclidean spinor Green's functions in the space-time of a straight cosmic string,''
  J.\ Math.\ Phys.\  {\bf 36}, 3694 (1995)
  [gr-qc/9412050].

\bibitem{tpbh} 
  R.~Emparan,
  ``AdS / CFT duals of topological black holes and the entropy of zero energy states,''
  JHEP {\bf 9906}, 036 (1999)
  doi:10.1088/1126-6708/1999/06/036
  [hep-th/9906040].

\bibitem{chm}
H.~Casini, M.~Huerta and R.~C.~Myers,
  ``Towards a derivation of holographic entanglement entropy,''
  JHEP {\bf 1105}, 036 (2011)
  [arXiv:1102.0440 [hep-th]].

  
  \bibitem{Banados:1992wn} 
  M.~Banados, C.~Teitelboim and J.~Zanelli,
  ``The Black hole in three-dimensional space-time,''
  Phys.\ Rev.\ Lett.\  {\bf 69}, 1849 (1992)
  [hep-th/9204099].
  

  \bibitem{Kraus:2006wn} 
  P.~Kraus,
  ``Lectures on black holes and the AdS(3) / CFT(2) correspondence,''
  Lect.\ Notes Phys.\  {\bf 755}, 193 (2008)
  [hep-th/0609074].
  
  
  \bibitem{Caputa:2013eka}
  P.~Caputa, G.~Mandal and R.~Sinha,
  ``Dynamical entanglement entropy with angular momentum and U(1) charge,''
  JHEP {\bf 1311} (2013) 052
  doi:10.1007/JHEP11(2013)052
  [arXiv:1306.4974 [hep-th]].
  
   \bibitem{wip}
 P.~Caputa, M.~Nozaki and T.~Numasawa, work in progress.
  
  \bibitem{Hemming:2002kd} 
  S.~Hemming, E.~Keski-Vakkuri and P.~Kraus,
  ``Strings in the extended BTZ space-time,''
  JHEP {\bf 0210}, 006 (2002)
  [hep-th/0208003].




\end{thebibliography}
\end{document}